\documentclass[a4paper,fleqn,usenatbib]{mnras}

\usepackage{newtxtext,newtxmath}

\usepackage[T1]{fontenc}
\usepackage{ae,aecompl}
\usepackage{xcolor}
\usepackage{ulem}

\usepackage{subfig}
\usepackage[export]{adjustbox}

\usepackage{graphicx}	
\usepackage{amsmath}	
\usepackage{amssymb}	
\usepackage{verbatim}
\usepackage{bm}

\graphicspath{{./plots/}}

\newcommand{\teff}{$T_{\mathrm{eff}}$ }

\newcommand{\T}[1]{$T_{\mathrm{eff}} = #100\,{\mathrm{K}}$}
\newcommand{\Mwd}{$M_{\mathrm{WD}}$}
\newcommand{\MH}{$M_{\mathrm{H}}$}
\newcommand{\MHe}{$M_{\mathrm{He}}$}

\newcommand{\nHe}{$n_{\mathrm{He}}$}
\newcommand{\Pwd}{$P_{\mathrm{WD}}$}
\newcommand{\gaia}{{\it Gaia}}
\newcommand{\tbetween}[2]{{#1}~$\leqslant T_{\mathrm{eff}}/{\rm [K]} \leqslant $~{#2}}
\newcommand{\treverse}[2]{{#1}~$\geqslant T_{\mathrm{eff}}/{\rm [K]} \geqslant $~{#2}}

\newcommand{\mteff}{T_{\mathrm{eff}}}

\title[White Dwarf Spectral Evolution]{From Hydrogen to Helium: The Spectral Evolution of White Dwarfs as Evidence for Convective Mixing}

\author[T.~Cunningham et. al]{Tim Cunningham,$^{1}$\thanks{E-mail: t.cunningham@warwick.ac.uk}
Pier-Emmanuel Tremblay,$^{1}$ Nicola Pietro Gentile Fusillo,$^{1}$
\newauthor{Mark Hollands,$^{1}$ and Elena Cukanovaite$^{1}$}
\\
$^{1}$Department of Physics, University of Warwick, Coventry, CV4 7AL, UK
}

\date{Accepted XXX. Received YYY; in original form ZZZ}

\pubyear{2019}

\begin{document}
\label{firstpage}
\pagerange{\pageref{firstpage}--\pageref{lastpage}}
\maketitle

\begin{abstract}
We present a study of the hypothesis that white dwarfs undergo a spectral change from hydrogen- to helium-dominated atmospheres using a volume-limited photometric sample drawn from the {\it Gaia} DR2 catalogue, the Sloan Digital Sky Survey (SDSS) and the {\it Galaxy Evolution Explorer} ({\it GALEX}). We exploit the strength of the Balmer jump in hydrogen-atmosphere DA white dwarfs to separate them from helium-dominated objects in SDSS colour space. Across the effective temperature range from 20\,000~K to 9000~K we find that 22\% of white dwarfs will undergo a spectral change, with no spectral evolution being ruled out at 5$\sigma$. The most likely explanation is that the increase in He-rich objects is caused by the convective mixing of DA stars with thin hydrogen layers, in which helium is dredged up from deeper layers by a surface hydrogen convection zone. The rate of change in the fraction of He-rich objects as a function of temperature, coupled with a recent grid of 3D radiation-hydrodynamic simulations of convective DA white dwarfs - which include the full overshoot region - lead to a discussion on the distribution of total hydrogen mass in white dwarfs. We find that 60\% of white dwarfs must have a hydrogen mass larger than $M_{\rm H}/M_{\rm WD} = 10^{-10}$, another 25\% have masses in the range $M_{\rm H}/M_{\rm WD} = 10^{-14}-10^{-10}$, and 15\% have less hydrogen than $M_{\rm H}/M_{\rm WD} = 10^{-14}$. These results have implications for white dwarf asteroseismology, stellar evolution through the asymptotic giant branch (AGB) and accretion of planetesimals onto white dwarfs.
\end{abstract}




\section{Introduction}
\label{intro}

From models of stellar evolution it is generally considered that white dwarfs are born with canonical masses of hydrogen and helium, \MH/\Mwd\ $\approx 10^{-4}$ and \MHe/\Mwd\ $\approx 10^{-2}$, where \Mwd\ is the total mass of the white dwarf \citep{schoenberner83,iben1983,dantona90},  with more recent estimates providing a parameterisation as a function of stellar mass \citep{althaus15,romero17}. These values are determined based on nuclear burning rates following the main sequence lifetime of the progenitor to the white dwarf. Over some decades, however, analyses of pulsating white dwarfs and stars evolving through and beyond the AGB \citep{werner06,deGeronimo18}, as well as population studies of cool white dwarfs \citep{tremblay08,rolland18,blouin19}, have led to the understanding that this is in all likelihood an upper limit on the mass of hydrogen. There has been much interest in constraining the mass of light elements in white dwarfs over the last few decades to understand their formation and evolution. 

Pulsating DA stars (DAV or ZZ Cetis) provide the opportunity to probe the chemical structure of white dwarfs as they cool between $\approx$ 12\,500\,K and 10\,500\,K \citep{tremblay15b}. It is generally accepted that all (non-magnetic) DA stars will exhibit variability from non-radial g-mode pulsations as they move through this temperature range during their evolution \citep{brickhill83,brickhill91,bradley96,fontaine08}. One of the great benefits of studying this population of white dwarfs is that models of asteroseismology describing the oscillatory behaviour are most sensitive to, among other parameters, the mass of the hydrogen layer \citep{fontaine08}. Predicted pulsation periods typically decrease when the hydrogen layer mass is increased, with the mean period spacing also decreasing slightly \citep{bradley96}. Recent asteroseismological studies often require significantly smaller hydrogen masses than the canonical value ($\log M_{\mathrm{H}}/M_{\mathrm{WD}} \ll -4$) to allow matching observed and predicted g-mode pulsations \citep{giammichele16,romero17}.

The small hydrogen masses that are invoked to model observed pulsations in some ZZ Cetis are thought to be explained by late hydrogen burning during the AGB and post-AGB \citep{dantona90,herwig99,werner06,althaus10}. Numerical simulations and theoretical calculations have shown that thermal pulses during and shortly after the AGB phase are able to burn up almost all of the remaining hydrogen \citep{straniero02}.

The radii of DA white dwarfs derived using evolutionary models with thin hydrogen layers are smaller than those with thick layers for a given mass \citep{wood90,t17}. Overestimating the hydrogen layer thickness can also lead to ages that are up to $\approx$1~Gyr too old for the coolest known white dwarfs \citep{fontaine01}. The field of Galactic archaeology has found white dwarfs to be useful chronometers in studies of the solar neighbourhood \citep{tremblay14} and in determinations of the age of the Galactic disk \citep{winget87,fontaine01,wood90,leggett98,chen12} or the Galactic halo \citep{kalirai12,kilic19}. Providing an independent constraint on the occurrence of different hydrogen layer masses could help to improve the accuracy of these models. 

As white dwarfs cool, the total amount of hydrogen and helium present, either primordial or accreted, can influence their subsequent evolution, and in particular their spectral appearance. The study of the spectral evolution of white dwarfs \citep{sion84,fontaine87,bergeron01,tremblay08} thus provides a method to learn about their past history and internal structure. In this work we focus on the hydrogen content, although we note that the study of hot hydrogen-deficient PG1159 stars \citep{werner06,miller16}, carbon dredge-up in helium-rich atmosphere DQ white dwarfs \citep{pelletier86,coutu19,koester19}, as well as asteroseismology can also help to constrain the mass of helium.

After gravitational separation has occurred \citep{schatzman45}, it is thought that hot white dwarfs will either cool as helium-rich DO atmospheres with He {\small II} lines, or hydrogen-atmosphere DA white dwarfs with Balmer lines if there is enough hydrogen to float at the surface. In both cases, radiative levitation can still keep trace metal species for $\approx$100\,Myr, long after gravitational settling is complete \citep{chayer14,koester14,werner18}. Both types of white dwarfs can also develop convective instabilities in their atmospheres or envelopes as they grow older, allowing for further changes in spectral types.

If a white dwarf has a thick enough hydrogen envelope ($\log$~\MH/\Mwd\ $\gtrapprox -14$; \citealt{rolland18}; \citealt{genest-Beaulieu19}) the convection zone will initially be confined to the hydrogen atmosphere. Recent numerical simulations have constrained the onset of convection in DA white dwarfs to arise at 18\,000--18\,250~K \citep{cunningham19} which sets an upper limit on the temperature range over which convection can impact the evolution of these objects. As the convection zone grows with decreasing surface temperature, if the hydrogen layer is sufficiently small ($\log M_{\mathrm{H}}/M_{\rm WD} \lessapprox -6$), eventually the convection zone will reach the deeper helium layer \citep{fontaine01}. At this point the significantly larger reservoir of helium ($\log M_{\mathrm{He}}/M_{\rm WD} \approx -2$) is expected to be immediately mixed into the surface convection zone in a runaway process resulting in a larger helium-dominated convection zone, a process named {\it convective mixing} \citep{strittmatter71,shipman72,baglin73,koester76}. With convective velocities reaching $v\approx$~1 km~s$^{-1}$ the chemical profile will almost instantaneously become homogeneously mixed \citep{elena1,cunningham19}. The result will appear to be a helium-rich atmosphere white dwarf (DB spectral type with He {\small I} lines or DC type with no lines) that may have detectable hydrogen \citep[DBA or DA spectral types;][]{rolland18}.

If a white dwarf has a thin enough total hydrogen mass ($\log$~\MH/\Mwd\ $\lessapprox -14$), a different evolutionary path is expected. Either the full evolution is in the form of a hydrogen-deficient PG1159, DO, DB, and then DC atmosphere \citep{genest-Beaulieu19}, or alternatively when the DO white dwarf reaches $T_{\rm eff} \approx$ 45\,000\,K it transforms into a DA with a very thin hydrogen atmosphere. Such a hydrogen layer is sufficiently small that the underlying helium layer is expected to become unstable to convection in the range 30\,000 $\gtrapprox T_{\rm eff}$/[K] $\gtrapprox$ 18\,000 \citep{macDonald91,rolland18,genest-Beaulieu19}. Convective overshoot is then expected to rapidly mix the top hydrogen layer with the underlying small helium convection zone \citep{cunningham19,elena2}, resulting in the so-called {\it convective dilution} process. The result is a DB or DBA white dwarf like in the convective mixing process described above, albeit with a different range of possible hydrogen abundances \citep{genest-Beaulieu19}.

In principle the hydrogen abundances in DBA stars could be used to reconstruct their past evolution, but this is not accounting for the fact that accretion of planetary debris can significantly impact their hydrogen content. In fact, several DBA white dwarfs have orders of magnitude more hydrogen than would be possible by the convective dilution or convective mixing scenarios, and it is thought that the accretion of water-rich asteroids is the most likely explanation for the hydrogen abundance in these objects \citep{farihi11,raddi15,ngf17}. The study of spectral evolution is clearly complex and involves many competing models that need to be constrained with well defined samples of the local white dwarf population.

Observational statistical studies on the number ratio of hydrogen- to helium-atmospheres as a function of temperature have mainly been carried out using spectroscopically identified samples with a magnitude limit \citep{bergeron97,bergeron01,tremblay08,blouin19,ourique19,genest-Beaulieu19} with the exception of \citet{limoges15} who relied on the volume-limited 40\,pc sample. In particular, we note that studying spectral evolution with the Sloan Digital Sky Survey spectroscopic sample, the largest known such sample for white dwarfs \citep{kepler19}, involves complex completeness corrections that are still not fully understood \citep{ngf15}. Photometrically selected or volume-complete samples have a strong advantage because the selection effects are better understood, especially in light of the recent \gaia\ Data Release 2 \citep{gaia2016,gaia-dr2}.

In this work we make use of \gaia\ and a new robust photometric technique to study spectral evolution for volume-complete white dwarf samples. We utilise a catalogue of $\approx$ 260\,000 high probability white dwarf candidates from \gaia\ \citep{ngf19} to select among them those with $ugriz$ photometry from the Sloan Digital Sky Survey (SDSS; \citealt{blanton17}). The SDSS $u-g$ colour is sensitive to the Balmer jump in the range 20\,000 $\gtrapprox T_{\rm eff}$/[K] $\gtrapprox$ 9000, allowing us to separate white dwarfs with hydrogen atmospheres (H-rich) from those with helium atmospheres (He-rich) without using the much more incomplete SDSS spectroscopic sample. Coupling photometric data from \gaia, SDSS, and {\it GALEX} with a grid of 3D radiation-hydrodynamic simulations of convective DA white dwarf atmospheres \citep{tremblay13c,cunningham19}, we study the scenario of convective mixing which is expected to happen within that temperature range. We investigate the mass distribution of hydrogen layers in white dwarfs with the highest precision to date, albeit within a limited mass range of $-14 \lessapprox \log$ \MH/\Mwd\ $\lessapprox -10$ given the $T_{\rm eff}$ range allowed to be studied with our technique.

We first describe our observed sample selection in Section~\ref{sec:sample}. Section \ref{sec:model_params} discusses the atmospheric models used to fit the photometric and astrometric data for the determination of effective temperature and stellar mass. Section \ref{sec:results} highlights the key results from the investigation and Section \ref{sec:discussion} contextualises the implications of our study.

\section{Photometric Sample}
\label{sec:sample}
For this investigation we chose a volume limited sample of high-confidence white dwarfs using the \gaia\ DR2 catalogue built by \citet{ngf19}. The selection criteria used were; a quality cut (\Pwd\ $\geq0.75$) which returned 262\,480 objects, a parallax cut (parallax $\geq 7.5$~mas) returning 35\,056 objects, an effective temperature range determined by fits to the \gaia\ parallax and photometry ($20000 \geq T_{\mathrm{eff}}/\mathrm{[K]} \geq 9000$) returning 6512 objects and a cross-match with the Sloan Digital Sky Survey (SDSS) photometry which returned a final sample size of 2207 objects. We emphasise that a large fraction of these objects do not have SDSS spectra. The cut in effective temperature is not essential at this stage, but included to indicate the size of the working sample. In Section~\ref{sec:results} the parallax cut is also relaxed to explore the results at greater distances and diagnose whether small number statistics can impact our results. All significant results, however, are borne from the sample with cuts detailed in the aforementioned.  

\begin{figure}
 \centering
 \subfloat{\includegraphics[width=1.0\columnwidth]{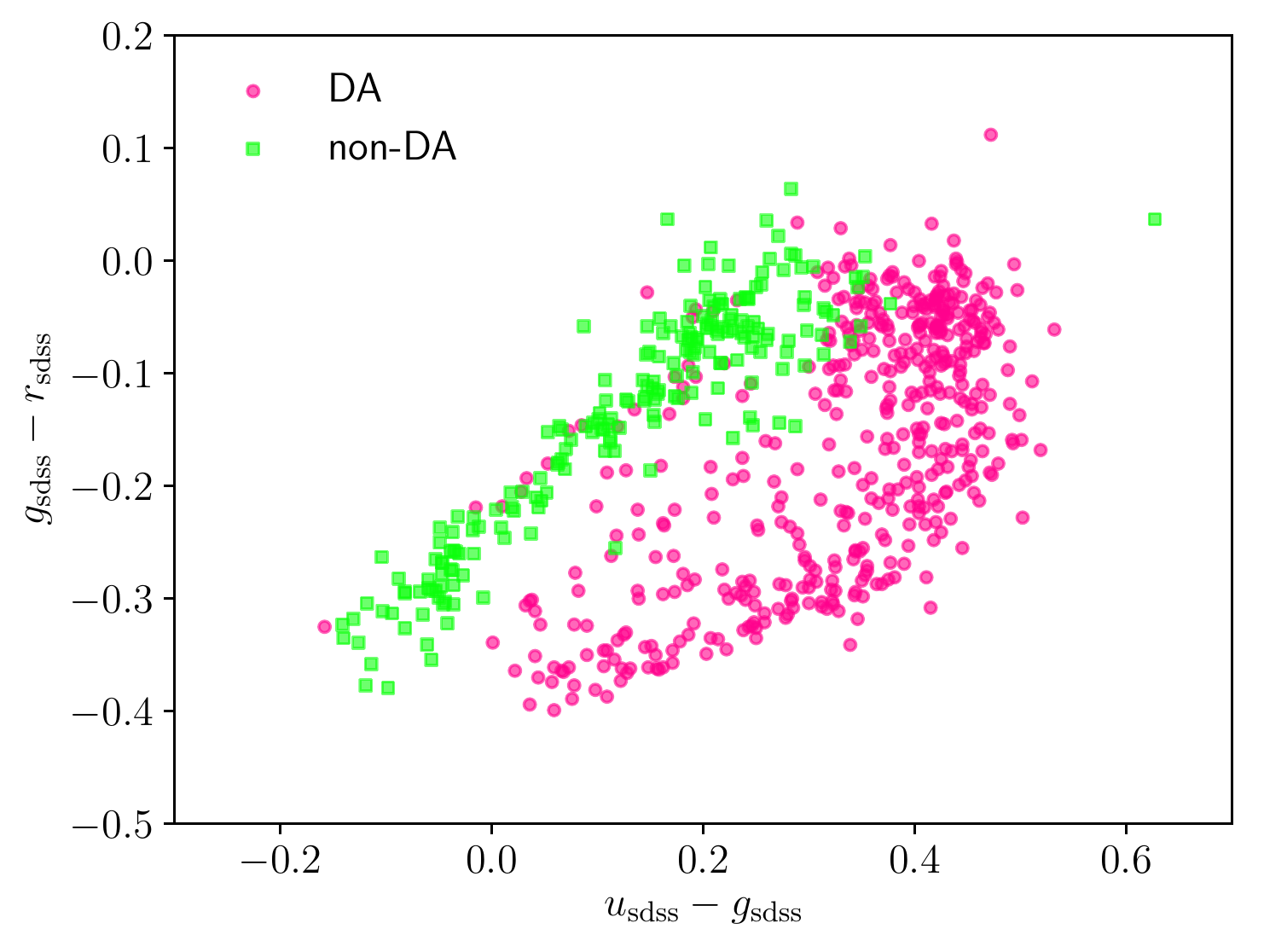}}
  \caption{Colour-colour diagram showing 690 objects from the volume limited white dwarf sample (parallax $\geq 7.5$~mas) with a SDSS spectral classification \citep{ngf19}. Of those shown, 479 are spectrally classified as DA (including DAH or DAZ) and 211 are classified as non-DA (including DB, DC, DQ, DBA, DAB and all magnetic and polluted variants of the aforementioned) with pink (circles) and green (squares), respectively.}
 \label{fg:col-SDSS-spec}
\end{figure}

In order to separate the hydrogen- and helium-dominated atmosphere white dwarfs from photometry alone we exploit the Balmer jump discontinuity which can be observed in spectra from sources with \teff$\approx$ 8\,000 -- 20\,000\,K. This discontinuity occurs at $\lambda \approx 364.4 - 380.0$~nm, depending on the stellar mass and strength of non-ideal effects \citep{hummer88}, making the SDSS $u$ and $g$ filters (central wavelengths of 354.3 and 477.0~nm, respectively) ideally suited to detect this feature. As a comparison colour we use $g-r$ with the central wavelength of $r$ at 623.1~nm. As an illustration, Fig.~\ref{fg:col-SDSS-spec} shows the sub-sample of 690 objects with SDSS spectra in a $u-g$, $g-r$ colour-colour plot where spectroscopically classified DAs are shown with pink circles and non-DAs are shown with green squares. The convention used throughout is that DAs include magnetic (DAH) and metal-rich (DAZ) objects, whilst non-DAs comprise DB, DBA, DAB, DC, DQ and all magnetic (H) and polluted (Z) variations therein.

\begin{figure}
 \centering
 \subfloat{\includegraphics[width=1.0\columnwidth]{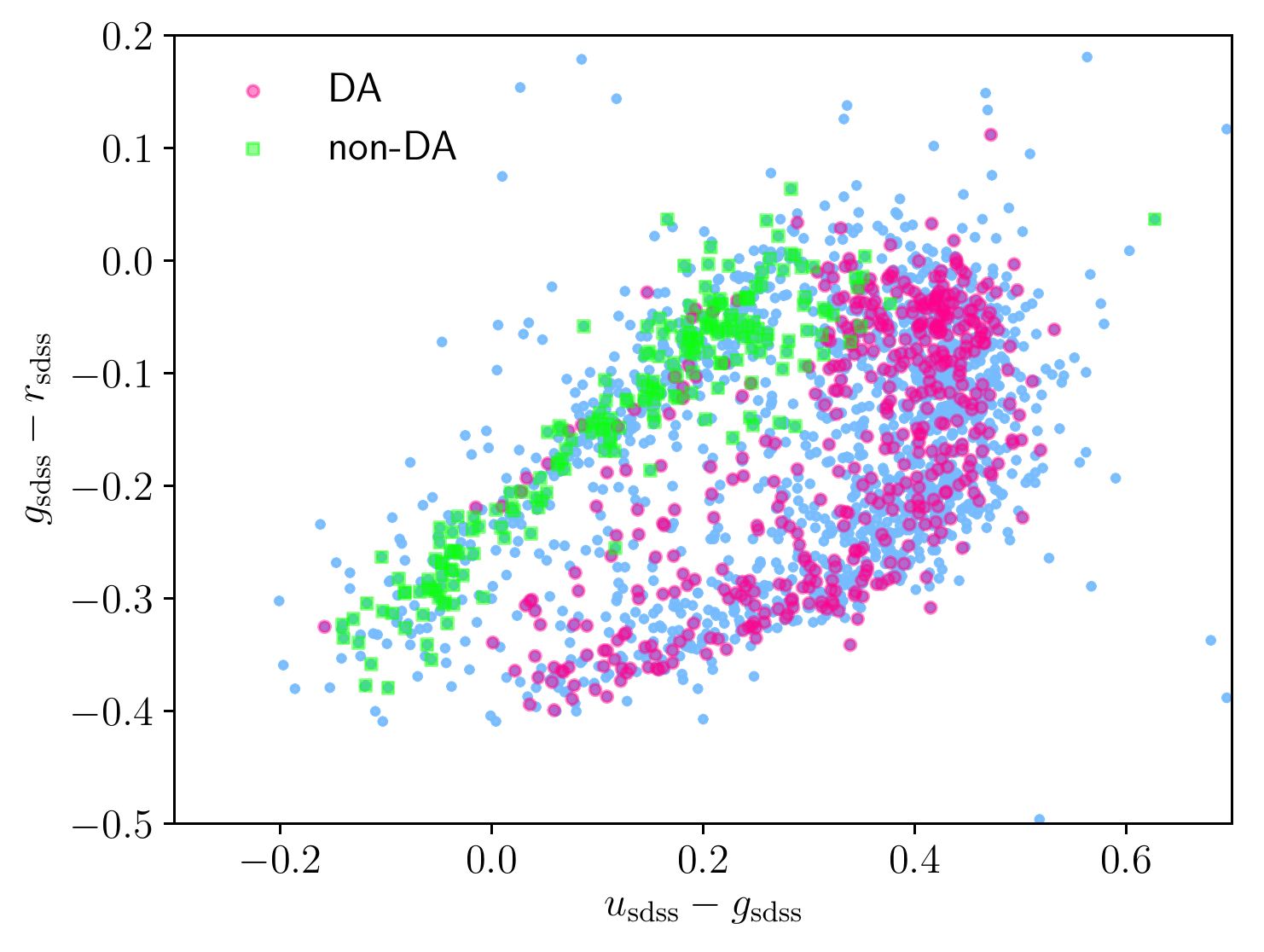}}
  \caption{Colour-colour diagram showing 2207 objects (blue) from the volume limited \gaia-SDSS photometric sample (parallax $\geq 7.5$~mas). Of those shown, 479 are spectrally classified as DA (including DAH or DAZ) and 211 are classified as non-DA (including DB, DC, DQ, DBA, DAB and all magnetic and polluted variants of the aforementioned) in pink (circles) and green (squares), respectively. 74 objects from the full sample populate a region of colour space outside the axis range shown. A manual inspection of the SDSS catalogue revealed that $\approx$65\% of those objects had been marked with a CLEAN=0 flag. The vast majority of all outliers are close to a bright star. We therefore remove these 74 objects from our sample.}
 \label{fg:col-SDSS}
\end{figure}

Fig.~\ref{fg:col-SDSS} shows the full sample from the photometric \gaia\ - SDSS cross-match in an analogous plot where objects with a spectral classification are shown in pink (DA, DAH, DAZ) and green (non-DA), while objects with blue points have no spectroscopic classification. 

Obtaining an immaculate separation of spectrally classified DA and non-DA objects is not feasible, with some DA objects (pink) occupying the same colour space as the strip of non-DA objects (green). An inspection of their spectra reveals that predominantly the DA type objects in the He-rich region of the colour plot are He-rich DA or DAZ \citep{zuckerman07,tremblay11,ngf17,rolland18}. In these instances - where there is sufficient helium to suppress the Balmer jump - these objects are correctly positioned photometrically, despite being classified as DA. That these objects sit in the photometric He-rich region is apposite for our analysis of the H- to He-rich atmosphere ratio. However there exists an area between the photometric clusters sparsely populated by objects which have an ambiguous atmospheric composition. To improve the separation we employ additional photometry from {\it GALEX} which we discuss in the following.

\subsection{{\it GALEX}}
\begin{figure}
 \centering
 \subfloat{\includegraphics[width=1.0\columnwidth]{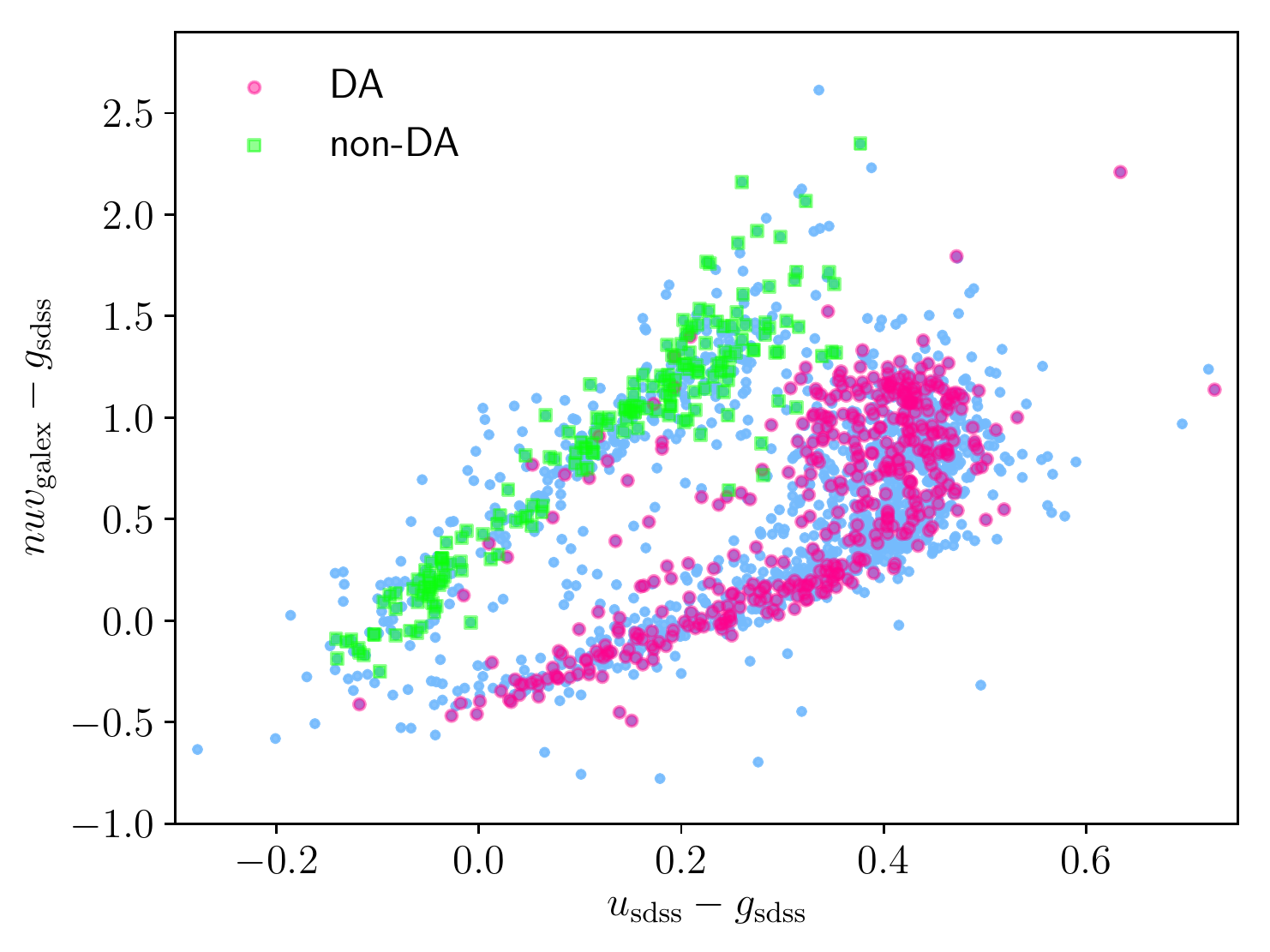}}
 \caption{Similar to Fig.~\ref{fg:col-SDSS} but utilising NUV from {\it GALEX}, which reduces the sample to 1781 objects (blue). Of those, 423 are spectrally classified as DA (pink circles) and 183 are classified as non-DA (green squares).}
 \label{fg:col-GALEX}
\end{figure}

The all-sky survey {\it GALEX} \citep{galex} provides photometry in the near- and far-ultraviolet for a large number of objects in our sample. We find that the separation between the spectrally classified sources (Fig.~\ref{fg:col-GALEX}) is increased when the colour $g-r$ (see Fig.~\ref{fg:col-SDSS}) is replaced by $nuv-g$. The large majority of DA stars found in the He-rich cluster in colour-space are He-rich DA white dwarfs, with only a handful of true contaminants (see Section~\ref{sec:results}). Approximately 400 objects in the sample were found to have unreliable or missing near-UV photometric data in {\it GALEX}, reducing the final working sample to 1781 high-confidence white dwarfs. Of those, 604 have SDSS spectral classifications as either DA[H,Z] (423) or non-DA (181). 

\subsection{Completeness}
\begin{figure}
 \centering
 \subfloat{\includegraphics[width=1.0\columnwidth]{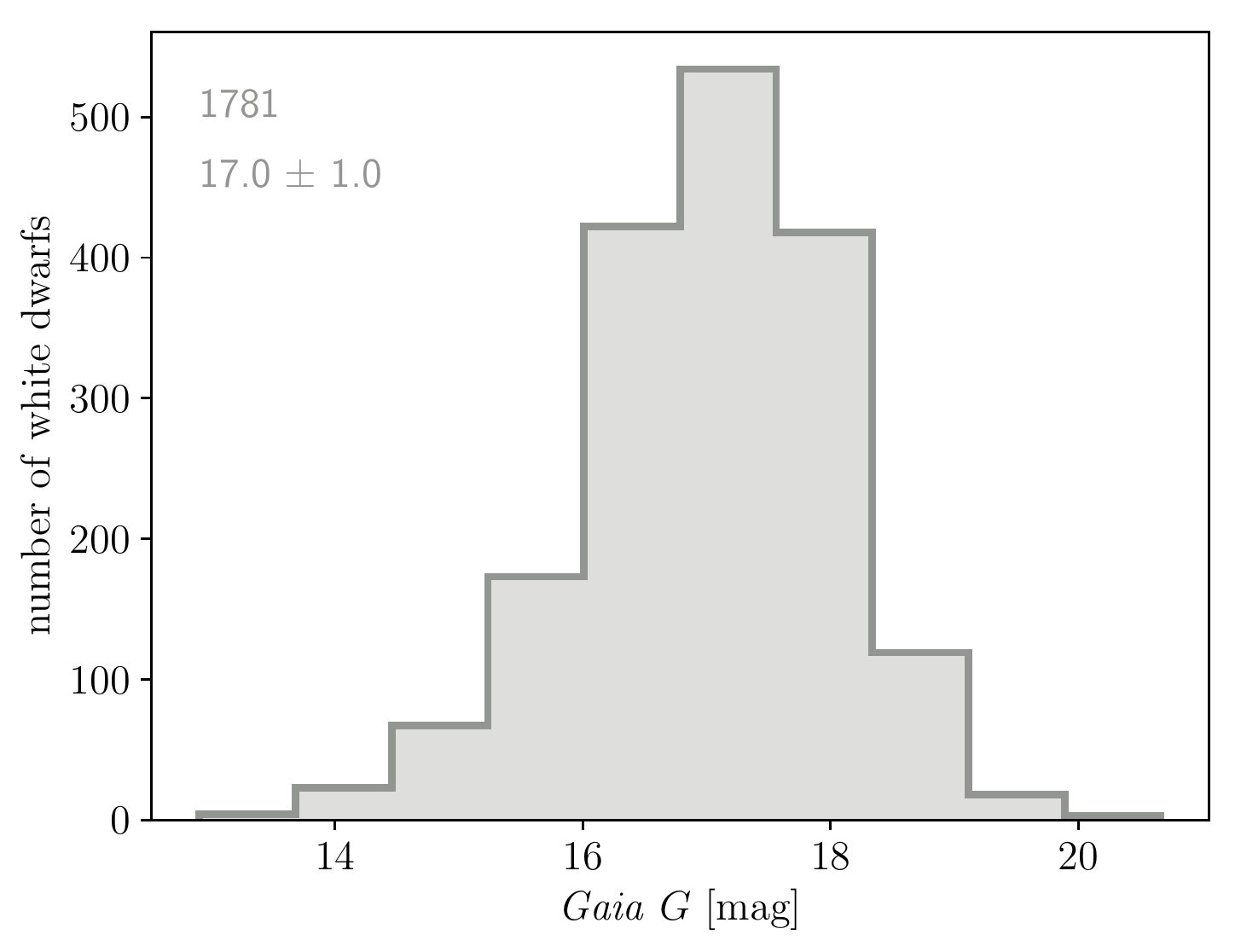}}
  \caption{Distribution of \gaia\ $G$ magnitudes for the 1781 objects included in our working sample after a cross-match with SDSS and {\it GALEX}. The mean magnitude is 17.0 with a standard deviation of 1.0.}
 \label{fg:mag-distr}
\end{figure}

\begin{figure}
 \centering
 \subfloat{\includegraphics[width=1.0\columnwidth]{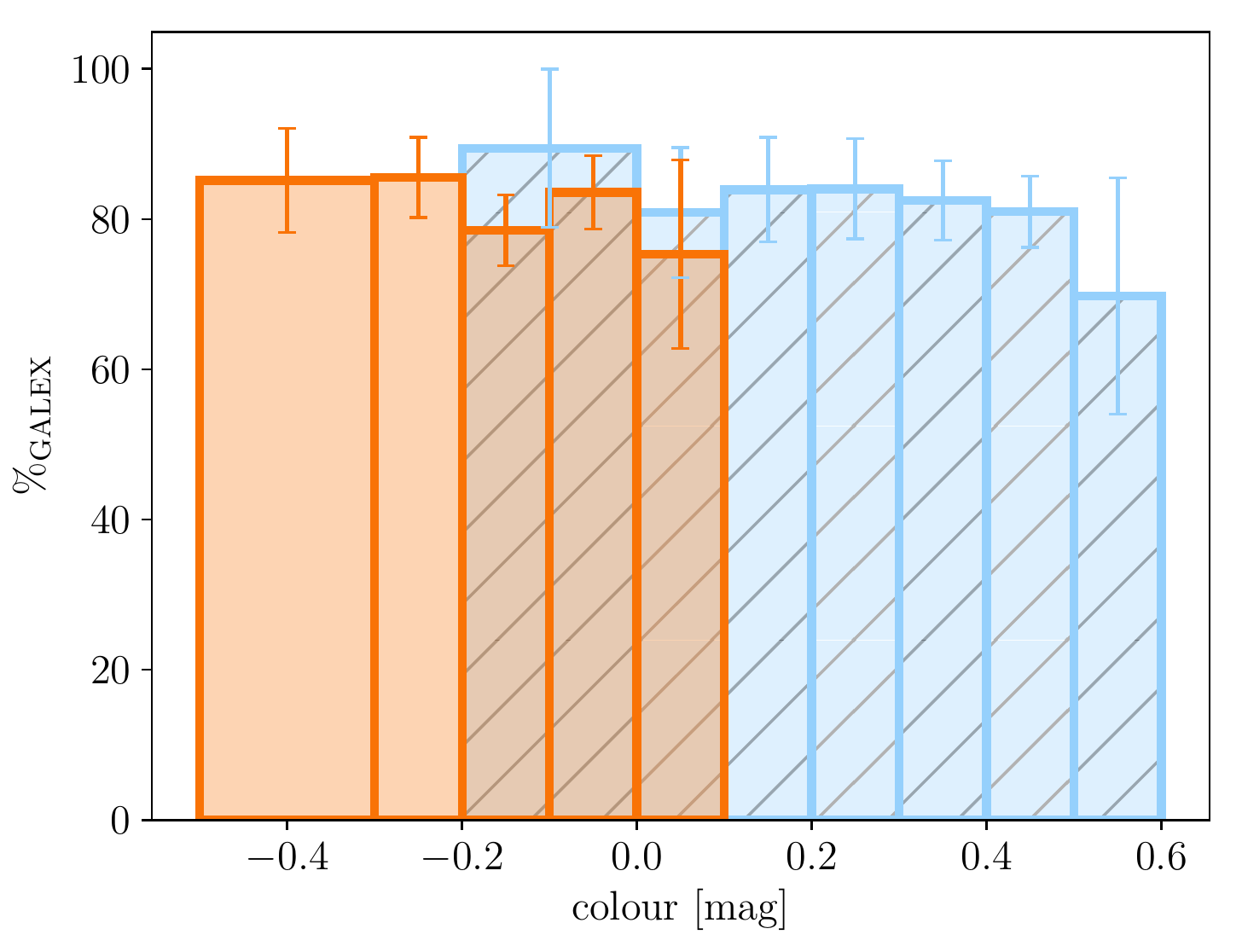}}
  \caption{Colour distributions of percentage of SDSS objects (Fig.~\ref{fg:col-SDSS}) recovered during the GALEX cross-match. In blue is the $u-g$ distribution for objects with colours from $-$0.2 to 0.6~mag. In orange we show the $g-r$ distribution for objects with colours from $-$0.5 to 0.1~mag. The colour ranges encompass the full extent of the locus of points in Fig.~\ref{fg:col-SDSS}.}
 \label{fg:col-distr}
\end{figure}

\begin{figure}
 \centering
 \subfloat{\includegraphics[width=1.0\columnwidth]{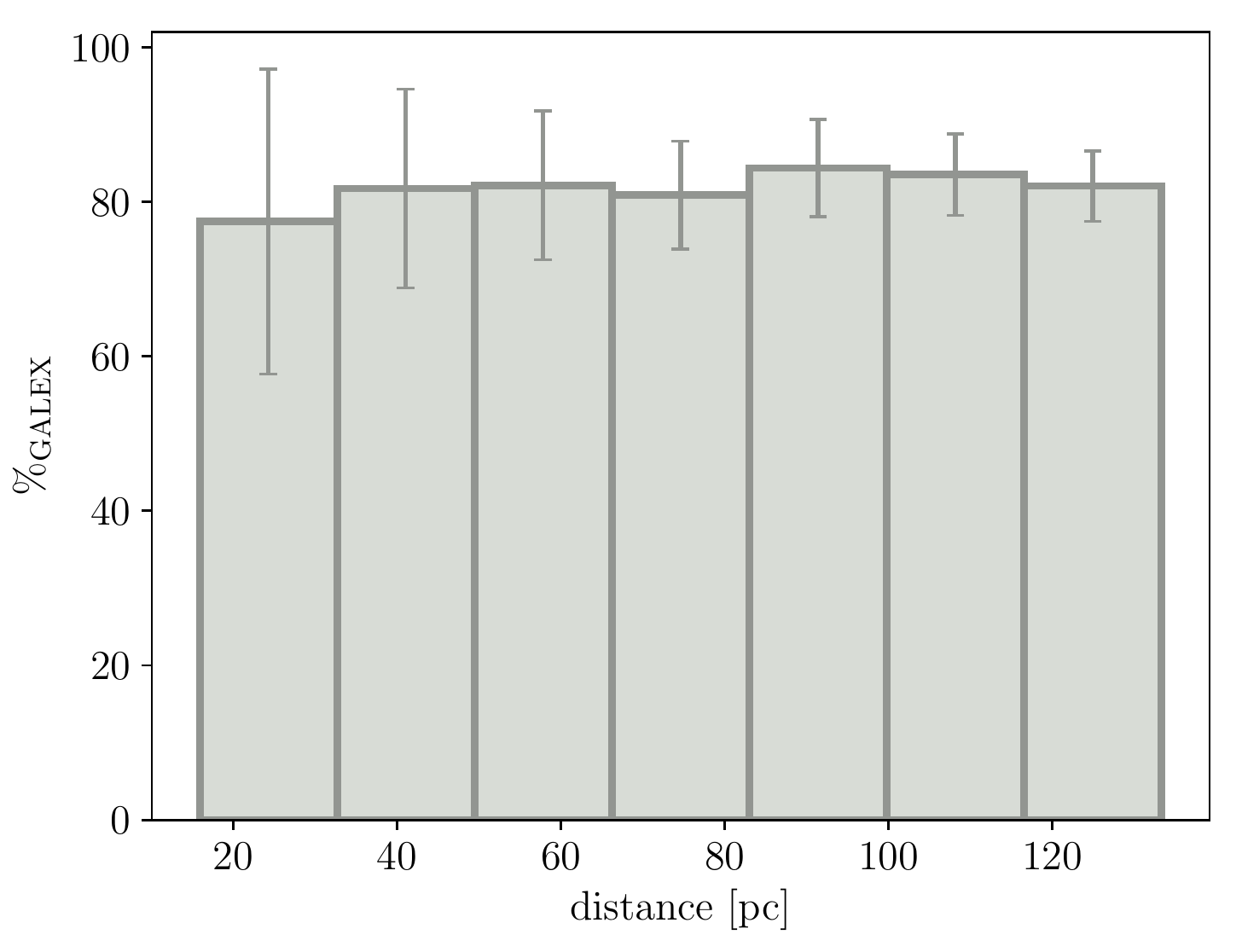}}
  \caption{Distance distribution, inferred from \gaia\ parallax, of the percentage of the SDSS sample (Fig.~\ref{fg:col-SDSS}) recovered during the {\it GALEX} cross-match.}
 \label{fg:dist-distr}
\end{figure}

We have shown that it is possible to calculate the fraction of H- and He-rich atmosphere white dwarfs relying solely on \gaia, SDSS and {\it GALEX} photometry for a limited $T_{\rm eff}$ range but it is also important to understand the completeness of this sample. A discussion on the completeness of the \gaia-SDSS photometric sample is in section 6 of \citet{ngf19} and summarised below. We emphasise that these authors also derive a \gaia-SDSS spectroscopic sample (their section 5), which is separate and not as complete since SDSS used complex surveying strategies to follow-up their photometric sources \citep[see, e.g.,][]{eisenstein06}. In Figs.\,\ref{fg:col-SDSS-spec}-\ref{fg:col-SDSS} and Section\,\ref{sec:results} we use this spectroscopic sample as a guide and comparison to our main photometric results. However, it is outside of the scope of this work to review the completeness of the spectroscopic \gaia-SDSS sample, which is known to vary considerably within colour-colour space, hence with $T_{\rm eff}$ and spectral type \citep{ngf15,ngf19}, and very likely impacting DA to non-DA ratios. Most earlier studies on spectral evolution have used such spectroscopic samples, although not as complete as with the selection defined in \citet{ngf19}. As a consequence we do not attempt to quantify the differences between these earlier studies and our method until spectroscopic completeness is better understood. Volume-complete spectroscopic samples, e.g. within 40\,pc \citep{limoges15}, still suffer from low number statistics in the $T_{\rm eff}$ range we are interested in.

In \citet{ngf19} the authors estimated the completeness of the \gaia-SDSS photometric cross-match based on the number of objects from the SDSS that were successfully retrieved by \gaia. They found that for white dwarfs with $G\leq20$ and \teff$\geq$ 7\,000~K, \gaia\ catalogued 60--85\% of the objects in the fairly complete SDSS footprint (sky images). However this includes white dwarfs at faint magnitudes and large distances that are too far to have a detectable parallax in \gaia. Given that our sample only includes objects within 133\,pc we expect the completeness of the cross-match to be much higher than this estimate. In fact \citet{hollands18b} find that the Gaia completeness is near 99\% for white dwarfs at 20\,pc, and \citet{ngf19} argue that there is no reason for this completeness to drop significantly within $\approx$ 100\,pc and for $G<20$~mag. Most importantly, \citet{ngf19} also quantified the completeness with respect to $u-g$ colour using the SDSS filters and found it to be colour independent (see their figure 18). This is a key parameter that makes our photometric method potentially more robust than earlier spectroscopic studies.

The SDSS footprint covers approximately one third of the sky meaning our sample is volume-limited only over the SDSS footprint. Within the distance set by the parallax cut our sample should be representative of the whole sky. The SDSS has a bright magnitude limit, with most white dwarfs brighter than $G \approx 15$ missing from the cross-match. Our sample is therefore not truly volume-limited. Our final magnitude distribution peaks at $G = 17.0$ with a standard deviation of 1.0 (Fig.~\ref{fg:mag-distr}) and, with 70 of the 1781 objects having $G<15$, we infer that we are likely missing no more than $\approx 4\%$ of all objects which is not expected to introduce a significant DA versus non-DA bias (see Section\,\ref{sec:PIER}).

The final sample uses a further cross-match with {\it GALEX}, and as such the completeness of {\it GALEX} is also important. To assess whether the inclusion of {\it GALEX} photometry introduces any bias within our sample we consider the colour distributions for the percentage of objects in the original SDSS sample which were also retrieved during the {\it GALEX} cross-match. Fig.~\ref{fg:col-distr} shows the distributions for $u-g$ (blue, hatched) and $g-r$ (orange, solid) colours for the final working sample. We find no significant colour dependence for objects found in the {\it GALEX} cross-match, where all bins are consistent with an $\approx$ 80\% retrieval rate to within 1$\sigma$.

We also show the distribution of objects recovered in the {\it GALEX}-SDSS cross-match as a function of distance (Fig.~\ref{fg:dist-distr}) to investigate whether any spatial bias could have been introduced. The distance is inferred using the \gaia\ parallax and we find the recovery rate of objects in {\it GALEX} shows no dependence on this parameter. As with the colour distributions we find all bins are consistent with a retrieval rate of 80\% to within 1$\sigma$. We conclude that the volume completeness of the final, working sample is likely to be 80\% that of the original SDSS sample and that no bias has been introduced as a result of including {\it GALEX} photometry. Given the SDSS bright magnitude limit and our \teff\ range, our {\it GALEX} sources do not suffer significantly from non-linearity problems \citep{camarota2014,wall2019}.

Finally we note that given our lower temperature limit of 9000\,K and the lower magnitude limits of \gaia, SDSS and {\it GALEX}, our distance limit of 133\,pc ensures that a negligible amount of white dwarfs are removed for being too faint in any of the surveys. At 9000\,K \gaia\ is the limiting survey and Fig.~\ref{fg:mag-distr} illustrates that our distribution peaks well above the lower magnitude limit of $G \approx 20$.

\begin{figure}
 \centering
 \subfloat{\includegraphics[width=1.0\columnwidth]{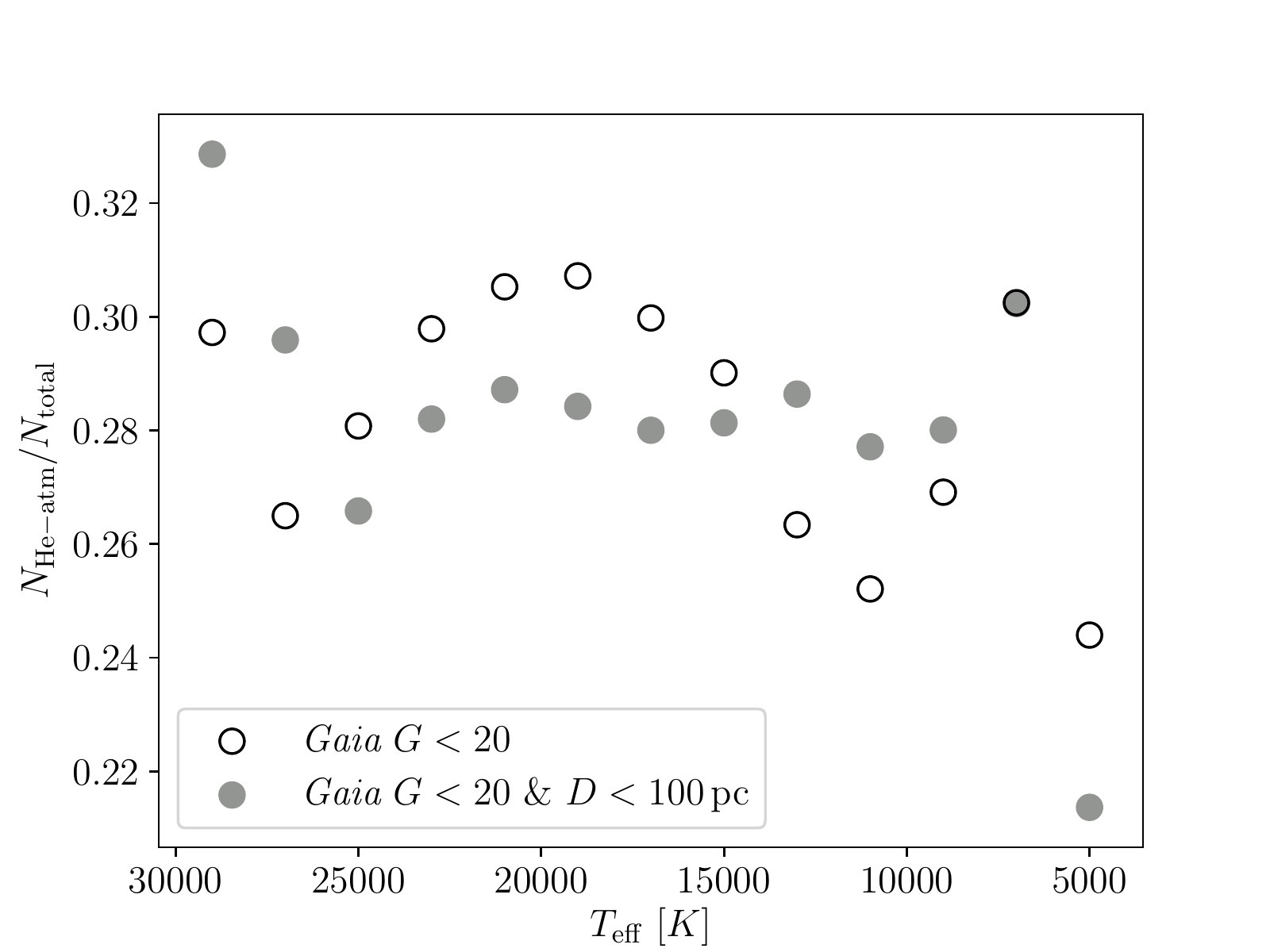}}
 \caption{Simulated temperature dependence of the helium dominated white dwarf fraction for volume (open circles) and magnitude limited samples (filled circles) assuming no spectral evolution at any time during the evolution. Input at birth included 72\% H-rich and 28\% He-rich atmospheres. The coolest bin is affected by the age of the Galactic disk and collision-induced opacities and should be interpreted with caution \citep{blouin19}.}
 \label{fg:PET-sims}
\end{figure}

\subsection{Absolute magnitudes and cooling rates}
\label{sec:PIER}

The use of a volume-limited sample largely removes possible selection biases due to absolute magnitude differences between spectral types but this concern still applies because of the bright magnitude limit of the SDSS. In addition a difference in the cooling rates of DA and DB stars could influence the ratio of spectral types as a function of temperature. To understand these biases we have simulated \gaia\ samples using the methods outlined in \citet{tremblay16} and \citet{ngf19}. These simulations assume a stellar formation rate \citep{tremblay14}, the Salpeter initial mass function, the \citet{cummings16} initial-to-final mass relation and the white dwarf evolution models of \citet{fontaine01}. Most importantly, 28\% of the white dwarfs are born with He-atmospheres (all with thin hydrogen layers), and 72\% with hydrogen atmospheres (where among those 14\% have thin hydrogen layers). In the subsequent evolution no spectral change is allowed. To eliminate random noise the white dwarf space density was artificially enhanced. Fig.~\ref{fg:PET-sims} demonstrates that for a volume limited sample, differences in cooling rates lead to small few percent-level changes of spectral type ratio over time. For reference, Fig.~\ref{fg:PET-sims} also shows the evolution for a magnitude-limited sample, where differences in absolute magnitude between DA and DB also play a role in the observed ratio. Once again, the effect from these biases is fairly minor, as outlined in \citet{tremblay08}. We conclude that any significant change ($>2\%$) in the observed $N_{\mathrm{He}}/N_{\mathrm{Tot}}$ ratio, in the range 20\,000 $\gtrapprox T_{\rm eff}$/[K] $\gtrapprox$ 9000 and for a volume-limited sample, must be caused by additional physical processes that happen during white dwarf evolution and that are not included in our simulations, such as convective mixing or accretion. 

\section{Atmospheric Composition}
\label{sec:model_params}

Using \gaia\ photometry and astrometry, most high-probability white dwarfs in the DR2 catalogue of \citet{ngf19} have a derived effective temperature from a dereddening procedure and model atmosphere calculation under the assumption of either a pure hydrogen or a pure helium atmosphere. The catalogue also makes use of the mass-radius relation of \citet{fontaine01} to derive a mass for each object.

The authors showed that, for a sample of 4778 bright DA stars, the \teff determinations using \gaia\ photometry were in agreement with those derived independently utilising photometry from SDSS and Pan-STARRS. 
 
\citet{tremblay19} and \citet{genest-beaulieu2019a} also made a comparison of photometric and spectroscopic effective temperatures derived from the SDSS. Individual objects were in agreement to within 1--2$\sigma$, but spectroscopic temperatures were systematically higher than those derived from \gaia\ photometry. It was concluded that this was most likely due to residual issues with the spectroscopic temperature scale.

\cite{bergeron19} have shown that using pure-He models in the photometric technique for objects below \teff $\approx$ 11\,000\,K results in a systematic effective temperature and mass offset compared to mixed H/He models (see their figures~10 \& 11). In the following section, the ratio of He- and H-rich objects - calculated from the SDSS-{\it GALEX}-\gaia\ photometric sample and empirical cuts described below - is computed using \teff bins of 1000~K. Hence the systematic offset in \teff and mass is not a significant concern for this analysis.

The cuts in $u-g$, $g-r$ and $u-g$, $nuv-g$ space employed to optimise the separation between the spectrally classified objects in Figs.\,\ref{fg:col-SDSS-spec}-\ref{fg:col-GALEX} are given, respectively, by the following equations:
\begin{align}
 (g-r) &= 0.8 \times (u-g) - 0.3 		\label{eq:phot_cut_g-r}		\\
 (nuv-g) &= 3.9 \times (u-g) - 0.1	 	\label{eq:phot_cut_nuv-g}
\end{align}

{\noindent}These empirical cuts are then used to estimate the ratio of H- to He-dominated atmospheres, below and above those lines, respectively. We then employ the \gaia\ photometric effective temperatures as described in \citet{ngf19} to transform this ratio into the context of white dwarf spectral evolution. White dwarf cooling takes place over a Gyr timescale \citep{dantona90} and employing evolutionary models would allow the study of spectral evolution as a function of age. For simplicity, effective temperatures are used as a proxy for age in this study. We use the pure-H and pure-He solutions for the H-rich and He-rich sides of our colours cuts, respectively. 
We note that the differences between H- and He-rich effective temperature using \gaia\ photometry are sufficiently small compared to the size of our bins that even if pure-H effective temperatures were used for all objects similar results would be obtained.

\section{Results}
\label{sec:results} 

\begin{figure*}
 \centering
 \subfloat{\includegraphics[width=1.0\columnwidth]{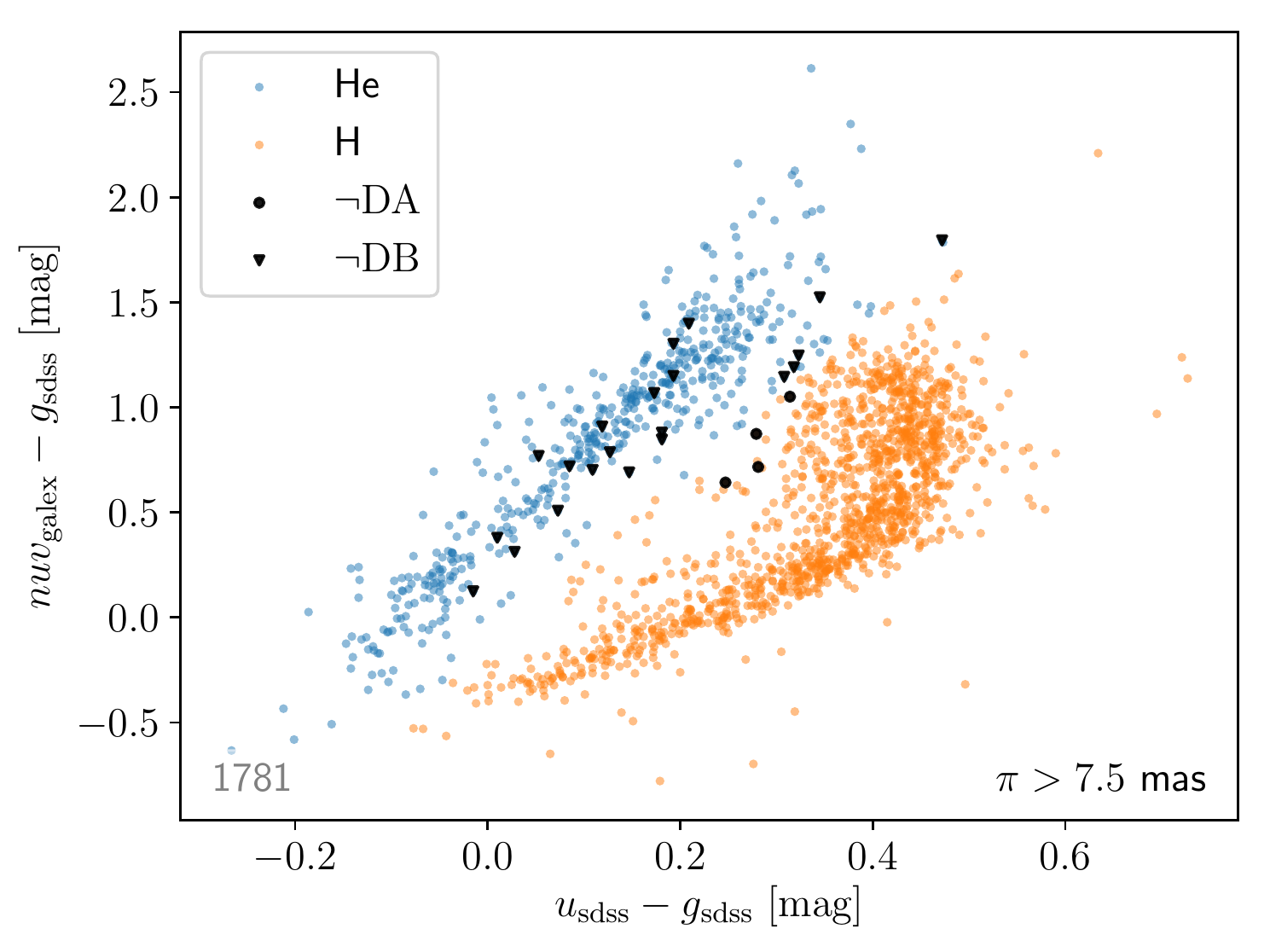}} 
 \subfloat{\includegraphics[width=1.0\columnwidth]{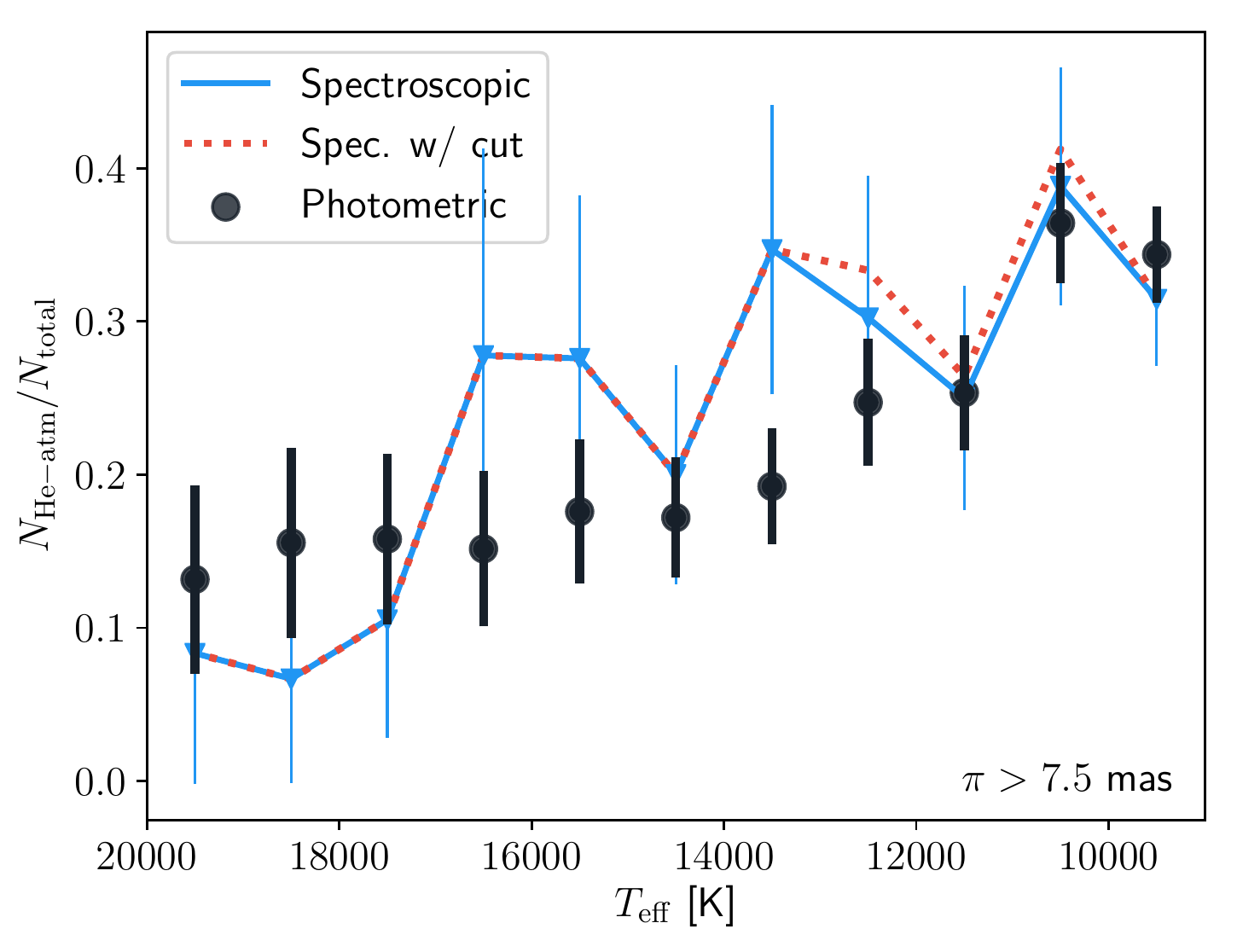}}\\
 \subfloat{\includegraphics[width=1.0\columnwidth]{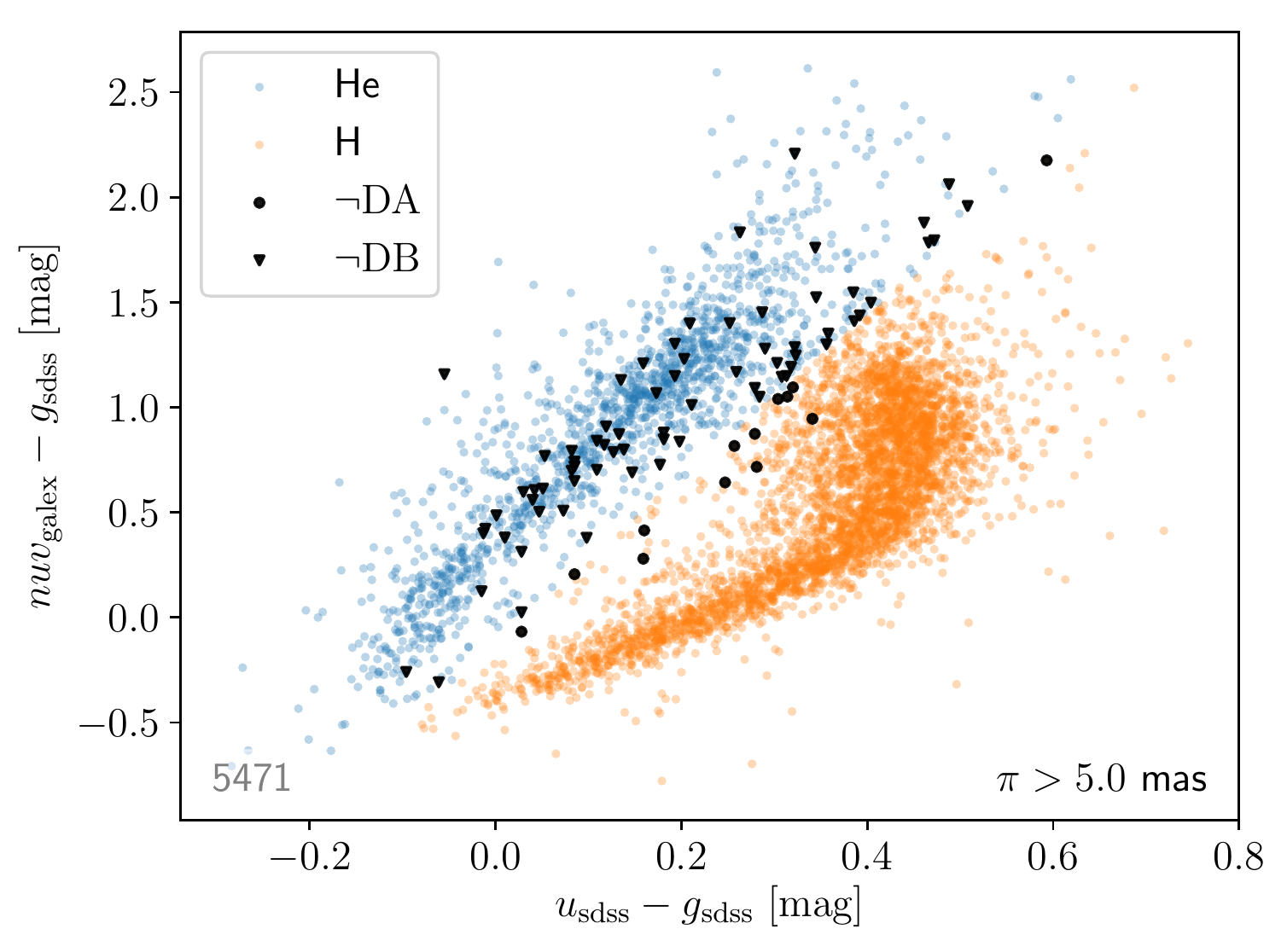}}
 \subfloat{\includegraphics[width=1.0\columnwidth]{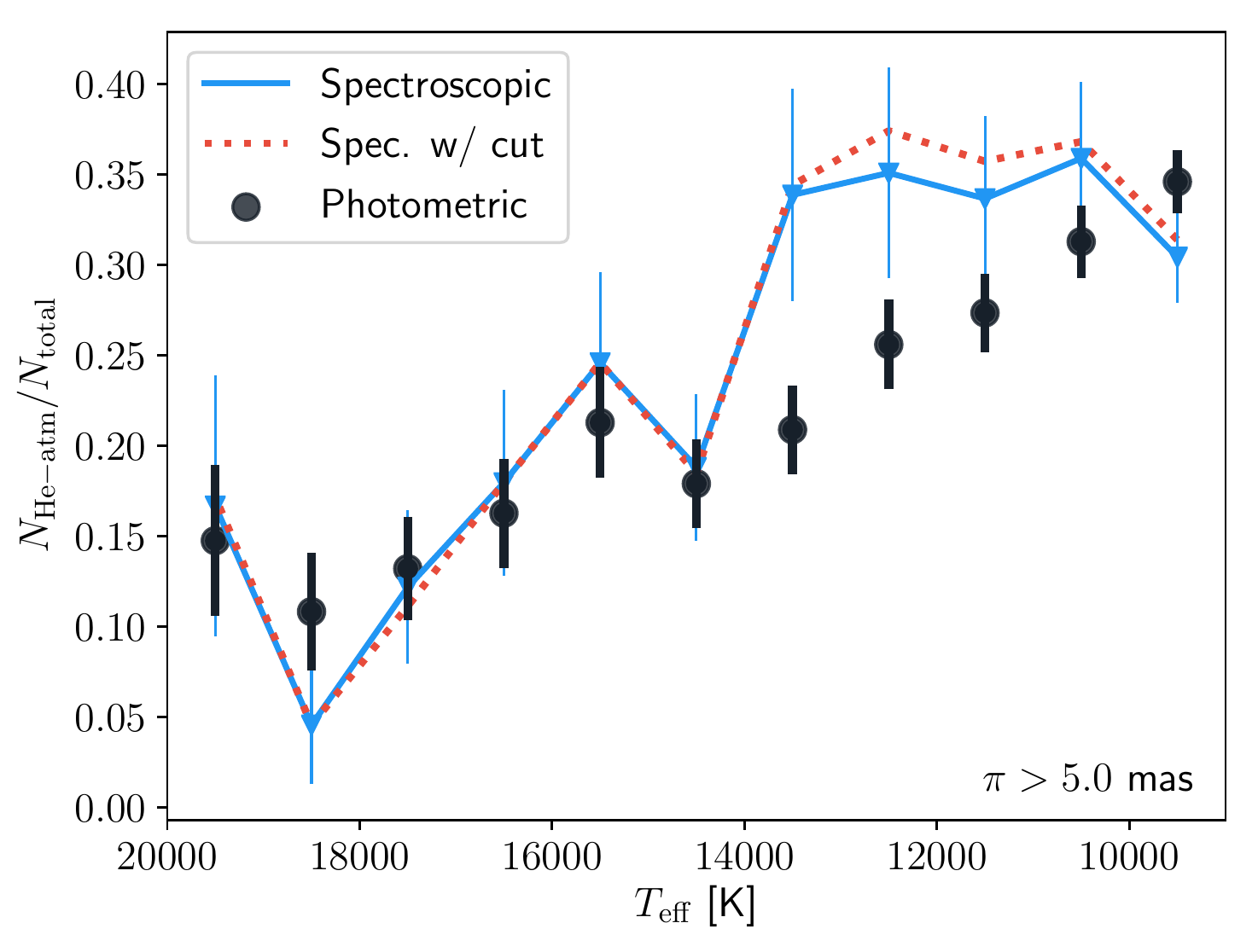}}\\ 
 \subfloat{\includegraphics[width=1.0\columnwidth]{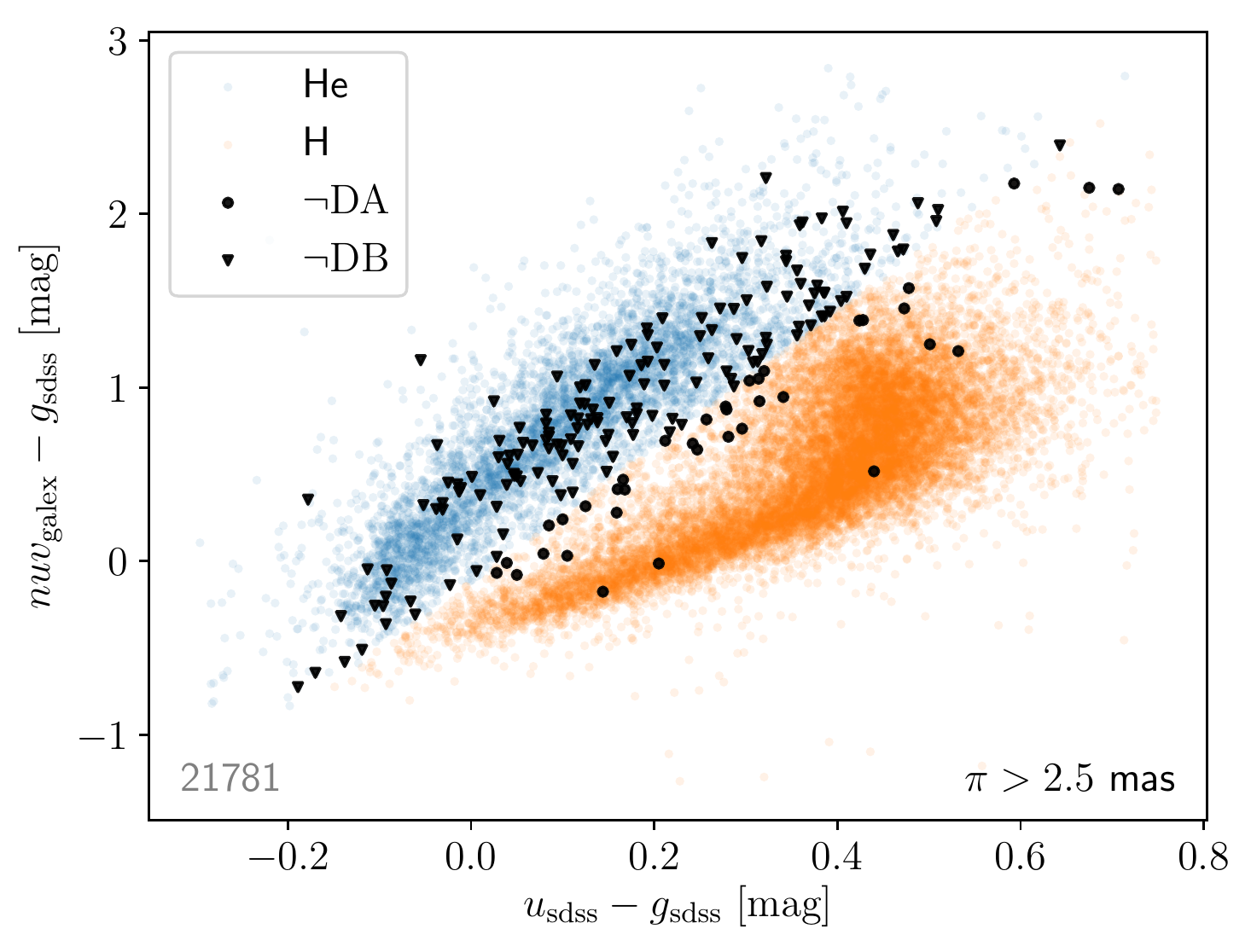}}
 \subfloat{\includegraphics[width=1.0\columnwidth]{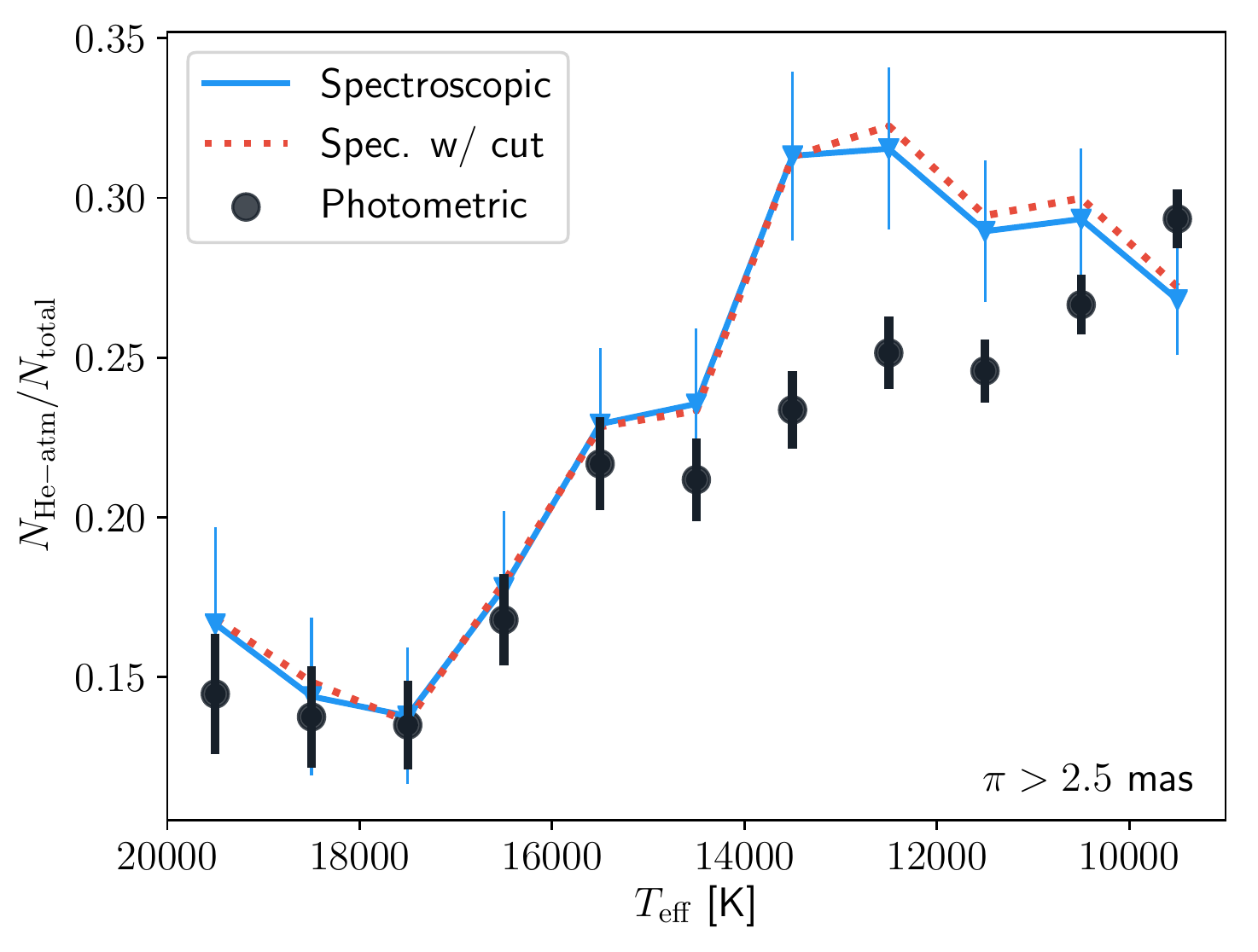}}\\ 
 \caption{{\it Top-to-bottom}: Same analysis shown for samples limited by increasing volume with the parallax (bottom right) and sample size (bottom left) indicated in the panels. {\it Left}: Photometric sample with separation of colour space for hydrogen dominated (orange) and helium dominated atmospheres (blue) according to Fig.~\ref{fg:col-GALEX} and Eq.\,(\ref{eq:phot_cut_nuv-g}). Spectrally classified objects are a small subsample of the photometric sample. Any objects with a spectral classification which apparently disagrees with its photometric assessment are shown in black for DAs with photometry predicting He-dominance (triangles) and non-DAs with photometry indicating H-dominance (circles). {\it Right}: Ratio of helium dominated objects to total as a function of effective temperature based on the photometric cuts (black circles). We also show the spectroscopic ratio in the same bins with (red, dotted) and without (blue, solid) respect for the photometric cut. Error bars result from the propagation of Poisson errors (see Eq.~\eqref{eq:num_err}).} 
 \label{fg:phot-cut}
\end{figure*}

For our derivation of the number of H- and He-atmospheres $(N_{\rm H-atm}$ and $N_{\rm He-atm})$ we devote our attention to the colour cut defined by Eq.~\eqref{eq:phot_cut_nuv-g} utilising the near-ultraviolet from {\it GALEX}, which creates a cleaner separation than Eq.~\eqref{eq:phot_cut_g-r}. The top-left panel of Fig.~\ref{fg:phot-cut} shows the effect of the photometric cut, with objects on the hydrogen side coloured orange and those on the helium side coloured blue. This includes objects with spectral classification. Shown in black are the spectrally classified objects which apparently fall on the {\it wrong} side of the photometric cut with DA-type objects on the helium side (triangles) and non-DAs on the hydrogen side (circles). Of the 604 spectrally classified objects we find 26 to be on the incorrect side of the cut. However, we point out that our goal is to separate H- and He-rich objects, rather than DA and non-DA.

We note the misidentification is skewed toward DA-type objects with 5.2\% (22/423) of these objects on the He-rich side of the cut, compared to 2.2\% (4/183) for the non-DA types. An inspection of the spectra for the DA-type objects in the He-rich region finds the majority (16/22) are either He-rich DAs or DA+DC binaries \citep{rolland18,kepler19}, suggesting those objects may sit correctly in the photometric analysis. A further 5 objects on both sides show evidence of metal pollution or strong magnetic fields. Calcium lines in the former case can reduce the {\it u}-band flux \citep{hollands17}, emulating the Balmer jump, and strong magnetic fields can significantly disrupt the Balmer jump. The remaining 5 objects have \teff$\approx 9000$\,K and higher-than-average surface gravities, where we expect the strength of the Balmer jump to be marginal. We conclude that H- and He-rich atmospheres separate relatively well under the photometric colour cut chosen, with 96--98\% of objects being assigned the correct composition.

The spectroscopic identifications serve as an indication of the photometric regions corresponding to hydrogen and helium dominance. Given a large enough and unbiased sample of spectra one could study the atmospheric composition of each object in turn and model the spectral evolution in that fashion. With the small size or biases present in the current spectroscopic samples this is not a well-justified approach, which is why we focus our analysis on the photometric sample.

The top-right panel of the figure shows the photometric fraction, $N_{\mathrm{He-atm}}/N_{\mathrm{total}}$, as a function of effective temperature in bins of 1000\,K. As a comparison the red dotted line shows the same quantity but only including objects with SDSS spectra, i.e. following the photometric cuts without regard to the actual spectral type. In this case we also rely on the photometric temperatures since some spectral types, e.g., DC, DZ, and DQ white dwarfs, have uncertain spectroscopic temperatures. Finally, this can be compared to the spectroscopic non-DA to total ratio in the same bins with the blue solid line. It demonstrates that the combined presence, as discussed above, of a few He-rich DA, strongly magnetic DAH, and DZ white dwarfs, only have a minor impact on our results.

The same analysis was carried out for increasing distance with a parallax cut of $\pi>7.5$ (upper), 5.0 (middle) and 2.5~mas (lower) also shown in the figure. As photometric precision and volume completeness decrease with distance, we expect that our colour cut of Eq.\,(\ref{eq:phot_cut_nuv-g}) becomes less reliable to select atmospheric composition, and as such we favour the smaller volume photometric results. In contrast, larger distances may be able to overcome low number statistics for spectroscopic ratios, without necessarily adding more contaminants. The largest volume results in an approach that is closer to those of \citet{genest-Beaulieu19} and \citet{ourique19}.

The errors associated with the photometric and spectroscopic ratios are derived from propagated Poisson errors by the expression 
\begin{equation}
 \alpha = \left(\left(\frac{n_{\mathrm{He}}+\alpha_{\mathrm{He}}}{n_{\mathrm{tot}}} - \frac{n_{\mathrm{He}}}{n_{\mathrm{tot}}}\right)^2 + \left(\frac{n_{\mathrm{He}}}{n_{\mathrm{tot}}+\alpha_{\mathrm{tot}}} - \frac{n_{\mathrm{He}}}{n_{\mathrm{tot}}}\right)^2\right)^{1/2}
\label{eq:num_err}
\end{equation}
where $n_{\rm tot}$ and \nHe\ are respectively the total number of objects and inferred numbers of helium dominated atmosphere white dwarfs from the photometry or spectroscopy and $\alpha_{\mathrm{X}} = \sqrt{n_{\mathrm{X}}}$ represents the counting (Poisson) error on both quantities. 

Biases in the SDSS spectroscopic follow-up selection function, which depends on the SDSS colours \citep{ngf15}, implies that the spectroscopic ratio may be less reliable. However a comparison of the photometric and spectroscopic ratios for $\pi > 7.5$~mas finds them to be in agreement to within 1$\sigma$ throughout the effective temperature range studied, with the exception of the bin centered at \teff$=$ 13\,500~K. At larger distances we observe a more prominent disagreement in the same temperature range.

The Poisson errors derived do not take into account the error on the effective temperature. We investigate the validity of the expressed error margin by calculating the sum of normal distributions in effective temperature for all objects in the top, left panel of Fig.~\ref{fg:phot-cut}. The normal distribution for species $X$ ($=$H or He), $P_X(T_i)$, assigned to data point, $i$, obeys
\begin{equation}
 P_X(T_i) = \frac{1}{\sqrt{2\pi}\sigma_i}\exp{\left(\frac{(T-T_i)^2}{2\sigma_i^2}\right)}
\end{equation}
where $T_i$ and $\sigma_i$ represent the modelled effective temperature and associated error. An expression of the continuous ratio of He-rich to total objects would then be computed as 
\begin{equation}
 \frac{N_{\mathrm{He-atm}}}{N_{\mathrm{total}}} = \sum_{i=1} \frac{P_{\mathrm{He}}(T_i)}{P_{\mathrm{H}}(T_i)+P_{\mathrm{He}}(T_i)}
\end{equation}
We find that all error bars from the original histogram bins are intersected by this function, with the exception of the bin centered at \T{10\,5}. We conclude that the errors attributed to the histogram bins from number statistics are a reasonable reflection of the uncertainty on the estimated photometric ratio.

\begin{figure}
 \centering
 \subfloat{\includegraphics[width=1.0\columnwidth]{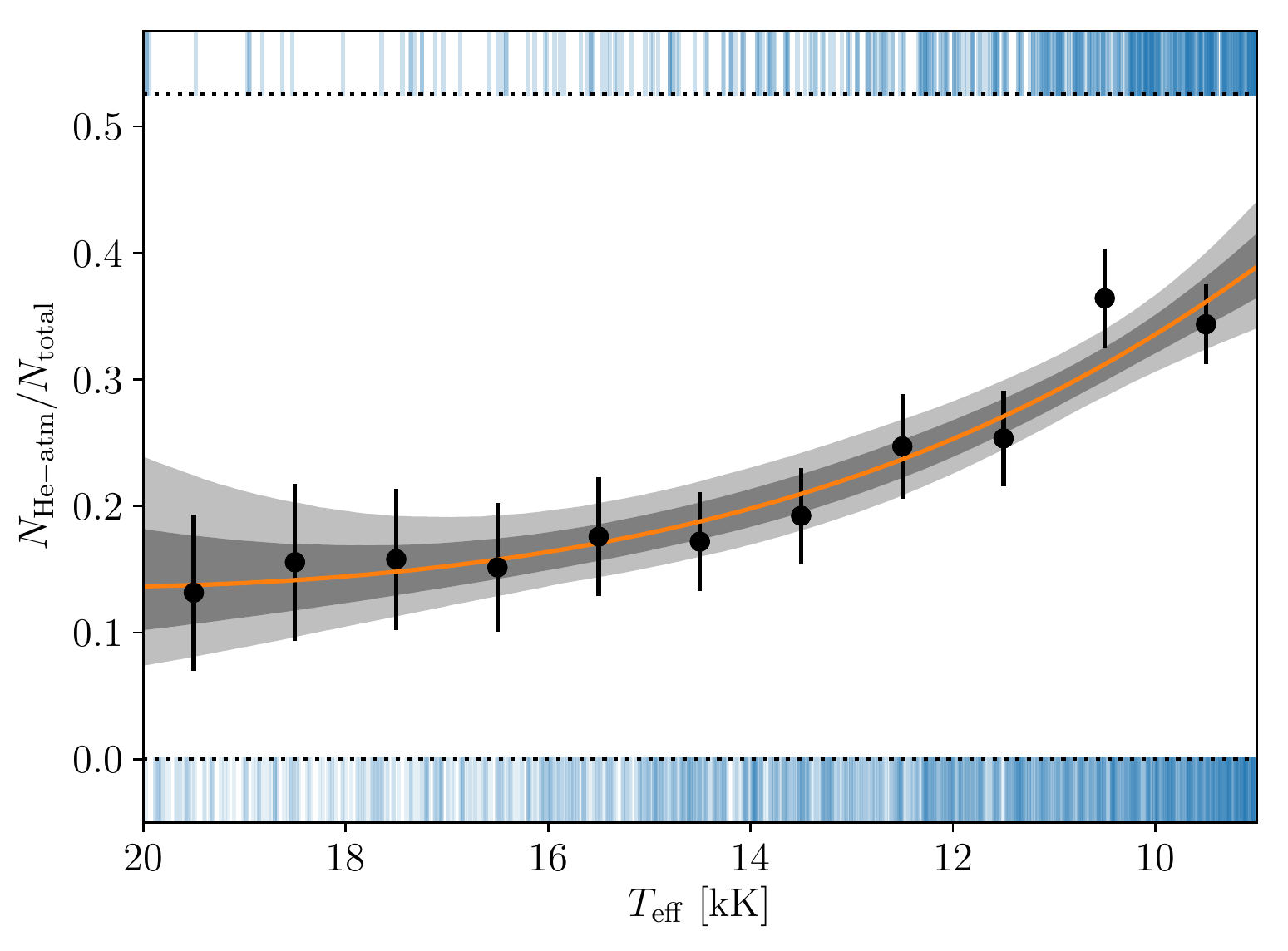}}
 \caption{Observed fraction of helium atmosphere white dwarfs derived by the method of logistic regression using Eq.~\eqref{eq:logit} and our $\pi > 7.5$ mas \gaia-SDSS-{\it GALEX} photometric sample. Filled error regions correspond to the 1$\sigma$ and 2$\sigma$ confidence intervals. Blue vertical lines indicate the effective temperature of a white dwarf with either a He-rich or H-rich atmosphere on the top and bottom, respectively. Photometric ratio derived using the binning method (see Fig.~\ref{fg:phot-cut}; top right) are also shown in black circles.} 
 \label{fg:continuous}
\end{figure}

\subsection{Fitting the spectral evolution}

One of the aims of this research is to model the observed spectral evolution. In order to do this we wish to express the fractional change of He-rich objects as a function of effective temperature. One could employ the values attributed to the histogram bins shown in Fig.~\ref{fg:phot-cut} (top-left panel), where the local gradient is the quantity of interest. However, the local gradient of the He-rich fraction versus effective temperature in the histogram picture has some dependence on the choice of histogram bins. This is not conducive to a robust result and so we instead seek a continuous function of effective temperature.

Logistic regression is a well established technique used in statistics and machine learning that estimates the likelihood of a function to describe a set of data, and it is increasingly being utilised in problems of an astrophysical nature \citep{hollands18b,chromey19}.
We assume that at a given effective temperature, $T_{\mathrm{eff}}$, the probability of a white dwarf being He-rich is given by $S(\bm{\theta}, T_{\mathrm{eff}})$, where $\theta$ is the vector of parameters that describe the form of function $S$ - whose shape we would like to find. Given that our data is categorical - objects are either deemed to be H-rich or He-rich - logistic regression is an apposite choice for finding this function. We summarise the methodology in the following. 

We define $S$ in terms of the logistic curve
\begin{equation}
 S(\bm{\theta}, T_{\mathrm{eff}}) = \frac{1}{1+\exp(-f(\bm{\theta}, T_{\mathrm{eff}}))}~~~,
 \label{eq:logit}
\end{equation}

{\noindent}where the function $f$ is defined as second-order polynomial in the natural logarithm of effective temperature such that
\begin{equation}
 f(\bm{\theta}, T_{\mathrm{eff}}) = \theta_0 \log \left(\frac{T_{\mathrm{eff}}}{[K]}\right)^2 + \theta_1 \log \left(\frac{T_{\mathrm{eff}}}{[K]}\right) + \theta_2~~~.
\end{equation}
Defining $S$ in this fashion ensures that the probability of an object being He-rich is confined to the interval [0,1]. This function differs from that used by \citet{hollands18b} with the addition of the second-order term in the polynomial. Adopting this term gives the model the freedom not to tend to zero, which is important as we expect the He-rich fraction to be non-zero for all temperatures. 

Now that we have a functional form with the freedom necessary to describe the likely shape of the probability of a white dwarf being He-rich, we invoke an optimization routine. The quantity we wish to maximize is the likelihood of the function $S$ with parameters $\bm{\theta}$ being the the best description of the data. Under the assumption that our data is independently Bernoulli distributed - that data has either the value 1, or 0 - the likelihood of a given $\bm{\theta}$ describing the data is expressed as
\begin{equation}
 L(\bm{\theta} | \mteff) = \prod_{i=1}^N{S_i^{y_i}(1-S_i)^{(1-y_i)}}
\end{equation}
where $N$ is the sample size and $S_i = S(\bm{\theta}, T_{\mathrm{eff,i}})$. The observation of whether an object is He-rich or H-rich is encapsulated with $y_i=1$ for He-rich objects or $y_i=0$ for H-rich objects. The optimization is made more straightforward by maximizing the natural logarithm of this quantity
\begin{equation}
 \log L(\bm{\theta} | \mteff) = \sum_{i=1}^N{\log(S_i)} + \sum_{i=1}^N{\log(1 - S_i)}
\end{equation}
Equivalently, one can minimize the negative log likelihood to find the best fit parameters. The optimization was performed using the Nelder-Mead algorithm built into the {\it minimize} function of {\sc scipy} which uses a simplex to find the minimum gradient in the parameter space \citep{scipy}.

The best fit parameters for $\theta_0$, $\theta_1$ and $\theta_2$ were $0.010\pm0.007$, $-0.43\pm0.20$ and $2.57\pm1.26$, respectively. The logistic curve with these parameters is shown in Fig.~\ref{fg:continuous} with the 1$\sigma$ and 2$\sigma$ confidence intervals shown in grey. The detections of He-rich or H-rich white dwarfs are shown in blue on the top and bottom axes, respectively. For comparison we show again the values from the histogram bin method presented earlier (Fig.~\ref{fg:phot-cut}) in black circles. We remind the reader that the logistic curve is not a fit to the black points, but derived independently as discussed via the method of logistic regression. On the statistical significance of the increase of the He-fraction, an inspection of the best-fit logistic function at 17\,000 and 9\,500~K finds that the increasing He-fraction across this temperature range can be considered a 5$\sigma$ result. 

Looking only at the histogram points, we find that for \treverse{20\,000}{14\,000} the percentage of helium dominated atmospheres lies between 10--20\%, with the general trend increasing by approximately 5\% toward the low temperature end of this interval, although we note that this is only a 1$\sigma$ result given the size of the error bars from Poisson errors. At lower effective temperatures, \treverse{13\,000}{10\,000}, we find a more significant increase, resulting in a final percentage of helium dominated atmospheres of 35--40\%.

\subsection{Mass distribution of thin hydrogen shells}

The hypothesis being tested is whether the inverse proportionality between effective temperature and the relative number of helium-atmosphere white dwarfs is due to convective mixing or convective dilution. 

{\it Convective mixing}, also referred to as dredge-up, is the process by which deeper material is dragged up by convective motions near the base of the (in this case hydrogen) convection zone. If the hydrogen convection zone is sufficiently close to the chemical boundary between the hydrogen and helium layers the material dragged up will be helium, and thus the two elements will mix. 

{\it Convective dilution} describes the interaction of the top of the helium convection zone with a thin hydrogen shell located above. As the upper boundary of the convection zone reaches the hydrogen layer the hydrogen will be steadily incorporated into the helium convection zone.

Our first hypothesis is that the observed change in He-rich fraction is caused by convective mixing alone. As the helium envelope is typically orders of magnitude more massive \citep{iben1983,romero19}, the prediction is that this runaway process quickly leaves a trace amount of hydrogen in a predominantly helium atmosphere \citep{rolland18}. Furthermore, convective motions in He-rich envelopes are many orders of magnitude faster \citep{fontaine1976} than any microscopic diffusion process that could separate helium and hydrogen \citep{koester09}, hence we assume this transition to be permanent.  In this picture we predict that the objects which change to appear helium dominated are white dwarfs with a total mass of hydrogen equal to the size of the convectively mixed region (i.e. convection zone size for chemical mixing) for a DA white dwarf at that temperature. Across the temperature range this allows us to predict the total hydrogen mass for a percentage of the total white dwarf population within 133~pc. For DA stars that do not mix, we can only estimate the minimum mass of hydrogen. 

\begin{figure}
 \centering
 \subfloat{\includegraphics[width=1.0\columnwidth]{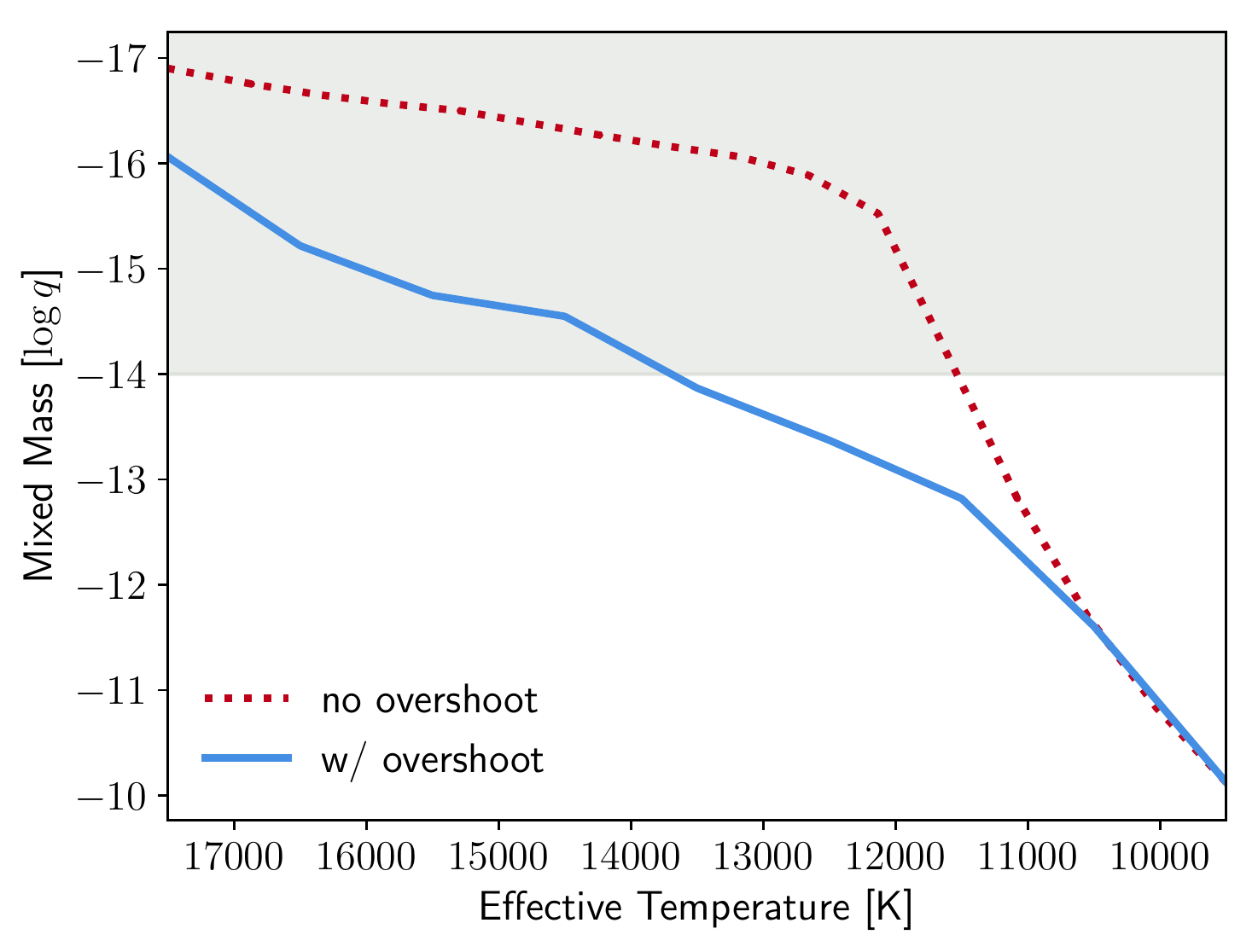}}
 \caption{Temperature dependence of the fractional mass ($q = M_{\rm H}/M_{\rm WD}$) of the convectively mixed region in a DA white dwarf. Results are shown for mixed masses derived from 3D simulations (\citealt{cunningham19}, blue solid) and 1D MLT (ML2/$\alpha=0.8$) with the convection zone defined by the Schwarschild criterion (\citealt{koester09}, red dotted).
 In this work we assume that when convective mixing occurs with the underlying helium layer, the mixed mass of hydrogen is equal to the total hydrogen mass in the star. The region in grey corresponds to hydrogen shell masses where convective dilution is expected to proceed \citep{genest-Beaulieu19}. }
 \label{fg:mixed-mass}
\end{figure}

To parameterize the convectively mixed region as a function of effective temperature we turn to previous results from 3D radiation-hydrodynamic simulations \citep{cunningham19}. Fig.~\ref{fg:mixed-mass} shows the predicted mass of hydrogen being mixed convectively. Results from 3D numerical simulations for white dwarfs in the temperature range \tbetween{11\,400}{18\,000} are shown in solid blue where the multi-dimensional treatment allows for the inclusion of convective overshoot without employing free-parameters. Mixing-length theory mixed masses taken from \citet{koester09}, with updated tables, are shown in dotted red. \citet{cunningham19} found that the mixed mass can be increased by 2.5 dex when convective overshoot is accounted for in these hydrogen atmosphere white dwarfs.

\begin{figure}
 \centering
 \subfloat{\includegraphics[width=1.0\columnwidth]{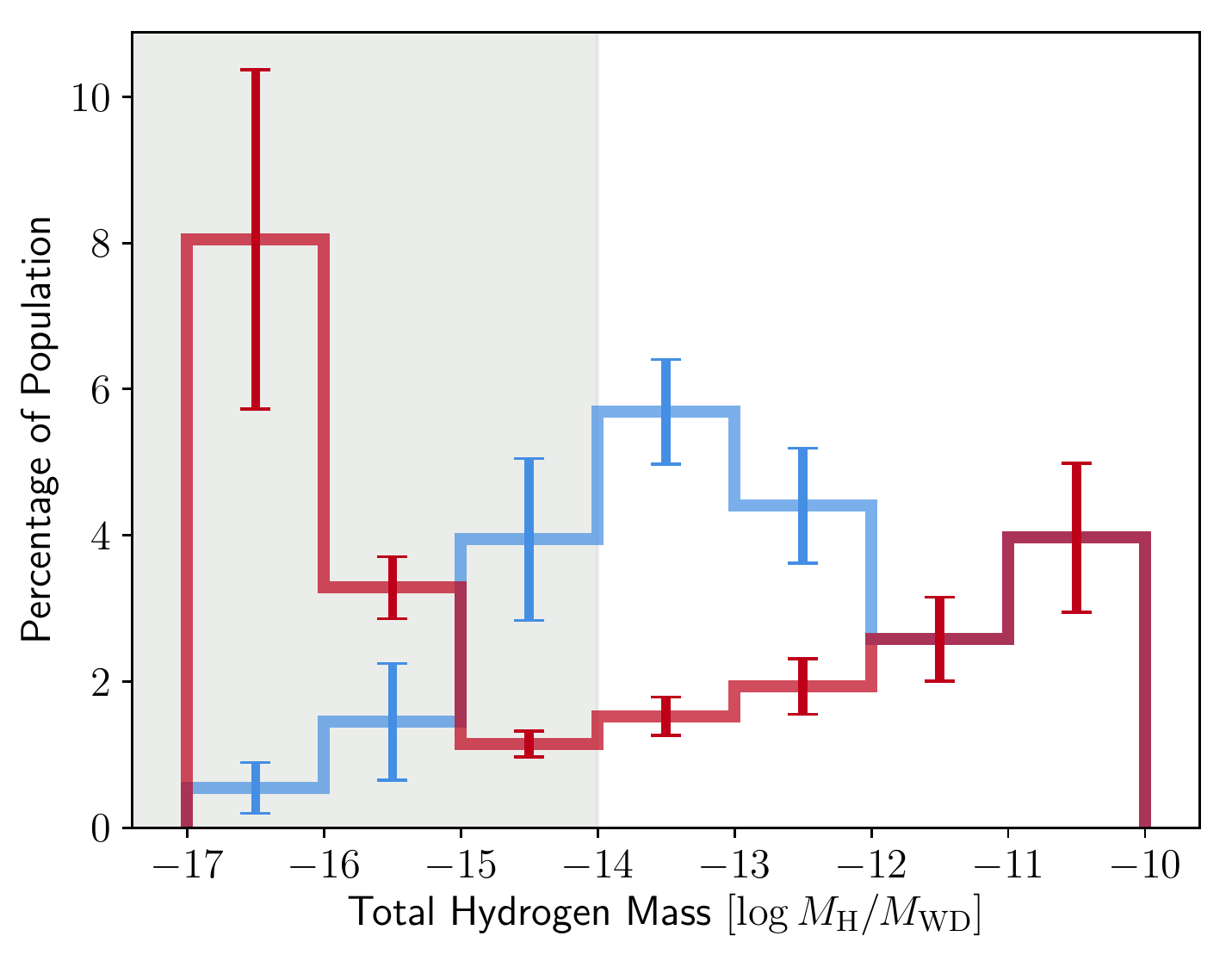}}
 \caption{Percentage of white dwarfs with a given total hydrogen mass inferred from the percentage change of He-dominated white dwarfs (see Fig.~\ref{fg:continuous}). Results are shown for the two mixed mass prescriptions described in Fig.~\ref{fg:mixed-mass} with 3D and 1D results in blue and red, respectively. The sum is less than 100\% because we do not include white dwarfs with lower or upper limits on the total hydrogen masses (see Section\,\ref{sec:discussion}). The region in grey corresponds to hydrogen shell masses that are {\it forbidden} for DA white dwarfs in our studied $T_{\rm eff}$ range according to convective dilution studies \citep{genest-Beaulieu19}. Under that scenario these objects would evolve as He-rich white dwarfs and show no spectral change in our studied $T_{\rm eff}$ range.}
 \label{fg:cum-mass}
\end{figure}

We show our derived total mass of hydrogen for white dwarfs within 133~pc in Fig.~\ref{fg:cum-mass} as a percentage of the total sample. Similarly to Fig.~\ref{fg:mixed-mass}, results are given for the convectively mixed mass with (blue) and without (red, dotted) the inclusion of convective overshoot. To derive the histogram values we sampled the parameter space of the function given in Eq.~\eqref{eq:logit} using normally distributed parameters with means and standard deviations equal to the best-fit parameters and their associated errors, respectively. The histogram values presented are the mean of the 10\,000 histograms calculated and the error is the standard deviation around that mean.

We find that including the larger mixed masses borne from the convective overshoot results increases the inferred total mass of hydrogen for $\approx$~15\% of the population by 2 dex. An increase in mixed mass due to convective overshoot is expected for \teff$<$~11\,400~K but is currently not accessible directly from 3D numerical simulations. As such no difference is accounted for in the current model for total hydrogen masses above $M_{\mathrm{H}}/M_{\mathrm{WD}} \approx 10^{-12}$ because of the convergence of 3D and 1D results at lower temperatures (see Fig.~\ref{fg:mixed-mass}). 

The second hypothesis includes the possibility of convective dilution occurring. We use results from \citet{genest-Beaulieu19} which predict that convective dilution will occur in white dwarfs with a thin hydrogen shell of mass of $\log M_{\mathrm{H}}/M_{\mathrm{WD}}<-14.0$ when they cool to \teff$\approx$ 25\,000~K. The inference is that these thinnest hydrogen shells are unlikely to exist by the time the white dwarf reaches \teff$<$ 20\,000~K. More massive hydrogen layers are predicted to suppress convection in the underlying helium layer, preventing convective dilution, and allowing the hydrogen and helium shells to remain stratified down to lower effective temperatures or until convective mixing happens. In this scenario convective mixing can therefore only occur in white dwarfs with total hydrogen masses $\log M_{\mathrm{H}}/M_{\mathrm{WD}}>-14$. Fig.~\ref{fg:mixed-mass} shows the region where convective dilution would have already happened in grey. In other words, convective mixing is only permitted for white dwarfs with \teff$<$ 13\,750~K in the 3D picture or \teff$<$ 11\,500~K in the 1D picture. 

In Fig.~\ref{fg:continuous} the temperature dependence of the helium-rich percentage appears to follow a steeper gradient for \teff$\lessapprox$ 14\,000~K, while a zero gradient could be used to fit the behaviour of the percentage for \teff$>$ 14\,000~K. This interpretation supports the hypothesis that convective dilution has already occurred in objects with $\log M_{\mathrm{H}}/M_{\mathrm{WD}}<-14.0$ by \teff$=$~20\,000~K and the next available mechanism for spectral evolution is the onset of convective mixing at \teff$\approx$ 14\,000~K. 

Fig.~\ref{fg:cum-mass} also includes the region of stratified hydrogen masses \textit{forbidden} by convective dilution in grey. It shows that, in this scenario, approximately half of the thin hydrogen shell masses predicted using 1D models of convection would be ruled out as nonphysical at the $T_{\rm eff}$ values of interest. So too would a quarter of the inferred hydrogen shell masses in the 3D picture. The fewer non-physical predictions of the latter model suggests the 3D overshoot picture provides a more robust description of convection in these objects. Furthermore, for the 3D results we find that the two bins associated with the smallest total hydrogen mass are consistent with zero within $2\sigma$. This provides evidence that the 3D overshoot model is in better agreement with the convective dilution model than a simple 1D treatment of convection alone. We note that the predictions of the convective dilution scenario are borne from a 1D mixing length treatment of convection. However, it was previously inferred that once the top of the helium layer becomes convectively unstable, a mechanism, that we suggest here is likely to be 3D convective overshoot, is able to dilute the top hydrogen layer into the helium convection zone \citep{rolland18}. As such 3D effects are already accounted implicitly in this scenario, although hydrodynamical simulations may give a more detailed picture of the onset of convective dilution, which is outside the scope of this work. 

We will now go on to discuss the implications of these results for the evolution of white dwarfs and time-dependent accretion.

\section{Discussion}
\label{sec:discussion}

We have presented a statistical analysis of a sample of 1781 white dwarfs to extract the most accurate characterisation of the fraction of He-rich white dwarfs as a function of temperature or cooling age. There is a highly statistically significant trend that the fraction of objects to have He-dominated atmospheres increases with cooling age. We find that this fraction increases from 10--20\% to 35--40\% between 20\,000 and 9000~K, corresponding to cooling ages ranging from 60 to 800 Myr. The implementation of a rigorous optimization method allowed to constrain this increase continuously across the temperature range. When combined with our convective mixing model, this provided a mass distribution of thin hydrogen shells (total hydrogen mass in the star) as a percentage of the total white dwarf population. We now consider some of the implications of these results outside of the white dwarf sample directly considered in this paper. 

\begin{figure}
 \centering
 \subfloat{\includegraphics[width=1.0\columnwidth]{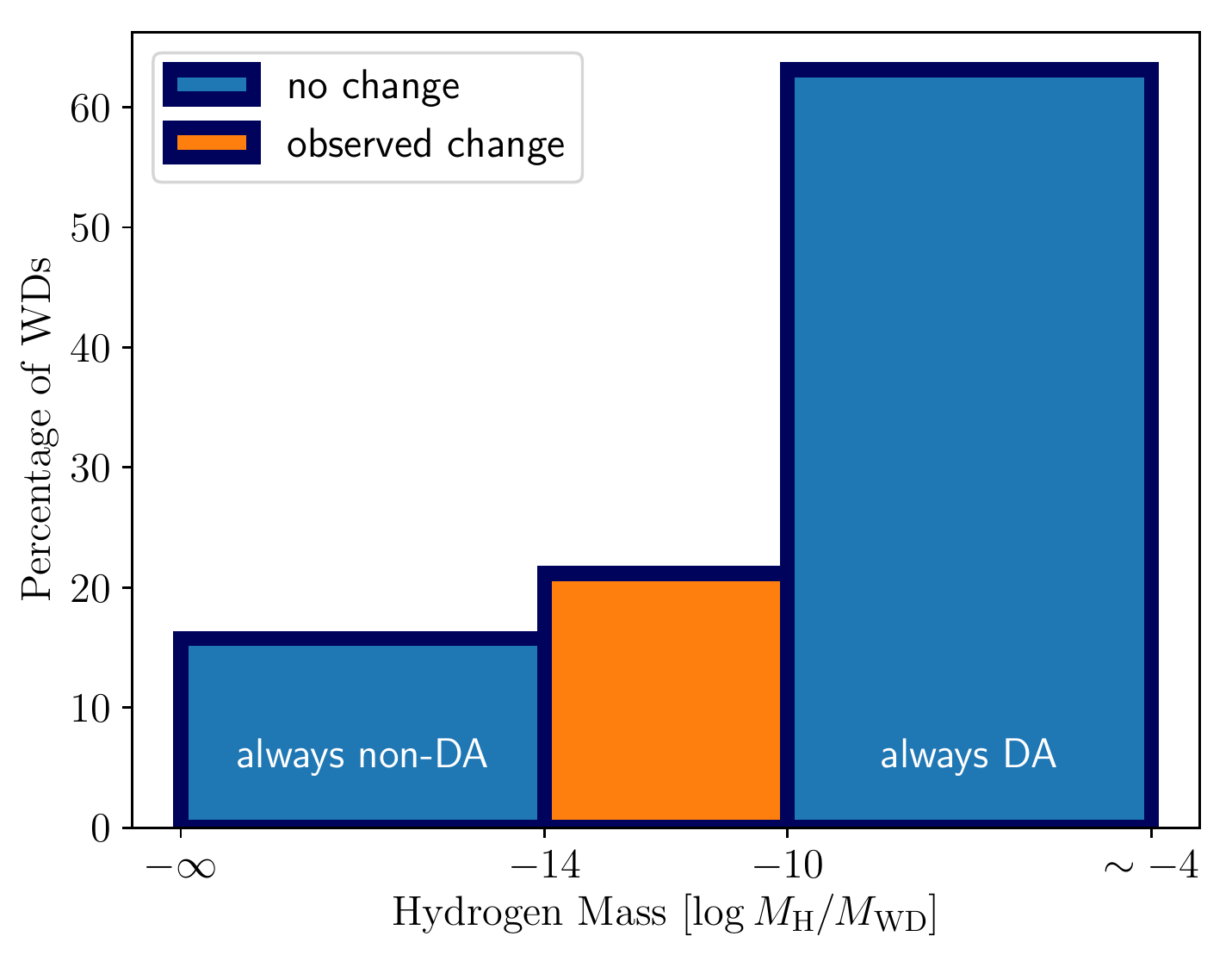}}
 \caption{Percentage of white dwarfs with a given total hydrogen mass. The percentage observed to undergo a spectral change from DA to non-DA within the range 20\,000 $\geqslant T_{\rm eff}/{\rm[K]} \geqslant$ 9000 are shown in orange. Those for which no change was observed within that range are shown in blue. Spectral evolution may be possible for temperatures outside of that range. }
 \label{fg:chunky-mass}
\end{figure}

The He-rich fractions observed at 20\,000\,K and 9000~K, respectively, allow for a mass distribution to be inferred across the full range of physically reasonable hydrogen shell masses. Fig.~\ref{fg:chunky-mass} shows this distribution for three bins. The central bin in orange comprises all objects observed to change from H- to He-rich in this study, where we use 3D convection for the hydrogen mass determination. The blue bins comprise all objects inferred to remain unchanged across the temperature range considered. We find that 61\% of white dwarfs must have a total hydrogen mass greater than $\log M_{\mathrm{H}}/M_{\mathrm{WD}}=-10$, and canonically it is considered that white dwarfs must have no more than $\log M_{\mathrm{H}}/M_{\mathrm{WD}} \approx -4$ \citep{romero19}. 

At the other extreme of masses we find that $\sim$15\% of white dwarfs must have a total hydrogen mass less than $\log M_{\mathrm{H}}/M_{\mathrm{WD}}=-14$. We note that this mass is dictated by the mass limit found in the convective dilution studies of \citet{rolland18} and \citet{genest-Beaulieu19}. These objects may have been DA stars at temperatures higher than 20\,000\,K. Our study can not derive the total number of white dwarfs that spend their full evolution as He-rich atmospheres. For this result, spectral evolution must be studied directly at hotter and cooler temperatures.

\subsection{Connecting with previous studies}
The temperature range considered throughout this study was defined by the region where the Balmer lines have the greatest prominence. For cooler temperatures this technique does not satisfactorily distinguish between H-rich and He-rich atmosphere white dwarfs. There have been recent efforts to quantify the ratio of DA to non-DA white dwarfs for cooler temperatures \citep{blouin19}.

We expect the contribution of convective mixing to the spectral evolution of white dwarfs to be less significant below 9000~K as the size of the convection zone becomes less sensitive to effective temperature. This was evidenced in \citet{blouin19} where they observed a small change in He-rich fraction for white dwarfs with effective temperatures 8000--4000~K. Therefore we would expect the He-rich fraction at the cooler end of our study to be somewhat comparable to the warmer end of the sample in their study.  In Table~\ref{tb:blouin-values} we show the fraction of He-rich white dwarfs between 8000 and 7000~K from literature values derived via various methods. To draw a comparison we consider the results obtained for the objects with effective temperature in the range 10\,000--9000~K. The histogram method returned a He-rich fraction of 0.34 $\pm$ 0.03, whilst the logistic regression method found a He-rich fraction of 0.36 $\pm$ 0.02 at 9500~K, i.e. in the centre of that temperature bin. We find our results to be in agreement within 1$\sigma$ with four earlier studies \citep{bergeron97,leggett98,bergeron01,tremblay08}. In contrast \citet{limoges15} and \citet{blouin19} find lower He-fractions. These differences could be investigated in the future with volume-complete spectroscopic follow-ups of local \gaia\ white dwarfs.

\begin{table}
	\centering
        \caption{Fraction of helium-rich atmosphere white dwarfs ($N_{\mathrm{He-atm}}/N_{\mathrm{tot}}$) in the temperature range 8000--7000~K presented in previous studies, collated by \citet{blouin19}.}
        \label{tab:grid}
        \begin{tabular}{lcc}
                Study & $N_{\mathrm{He-atm}}/N_{\mathrm{tot}}$\\ 
                \hline
                \cite{bergeron97} & $0.33 \pm 0.12$ \\
                \cite{leggett98}  & $0.45 \pm 0.23$ \\
                \cite{bergeron01} & $0.37 \pm 0.09$ \\
                \cite{tremblay08} & $0.31 \pm 0.06$ \\ 
                \cite{limoges15}  & $0.23 \pm 0.05$ \\
                \cite{blouin19}   & $0.14 \pm 0.03$ \\
                \hline
        \end{tabular}
        \label{tb:blouin-values}\\
\end{table}

\subsection{Hydrogen abundance in DBA white dwarfs}

\begin{figure}
 \centering
 \subfloat{\includegraphics[width=1.0\columnwidth]{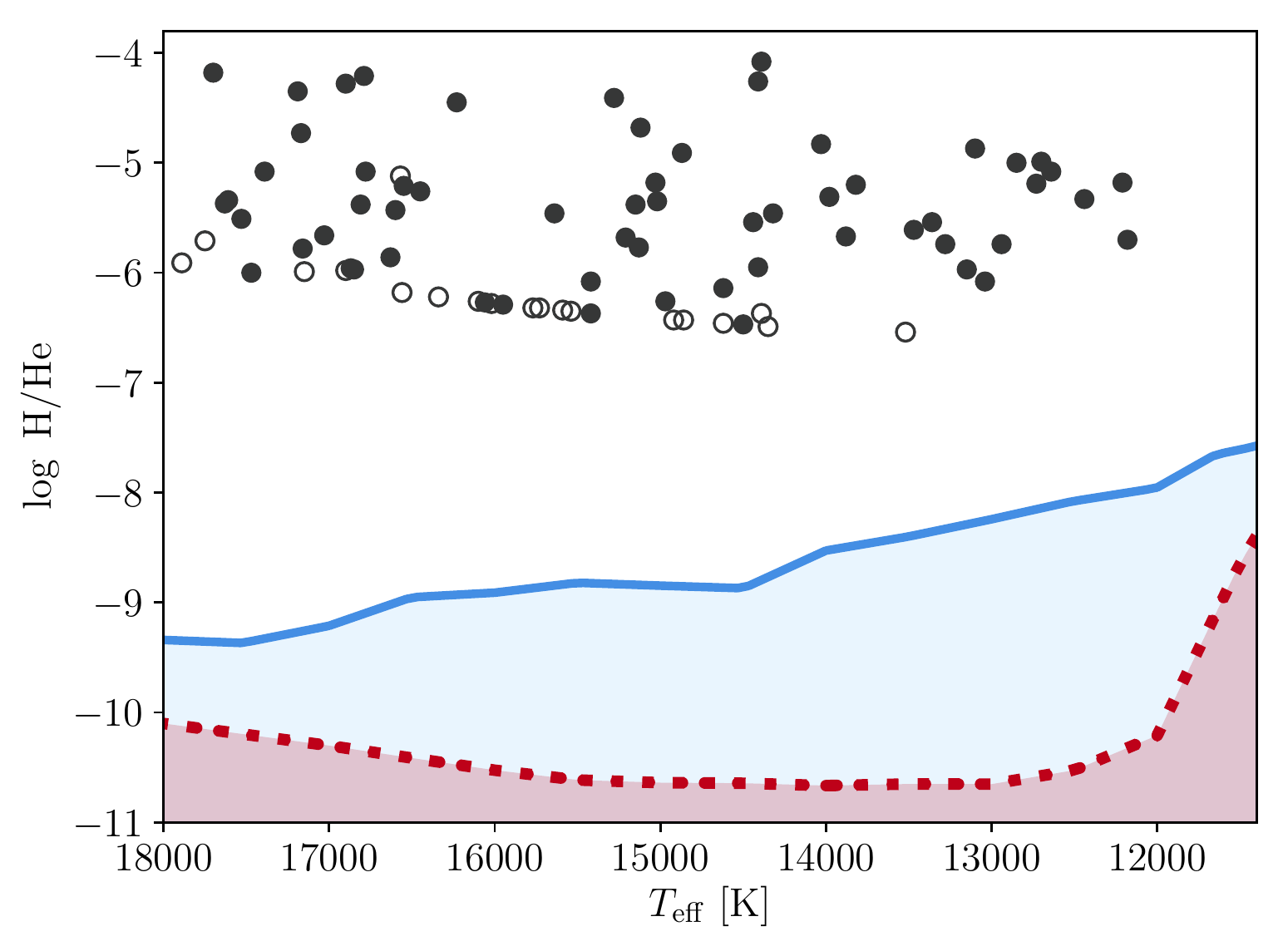}}
 \caption{Observed hydrogen abundances in DB(A) white dwarfs drawn from the spectroscopic observations of \citet{rolland18}. Filled circles show detections while open circles designate upper limits based on a non-detection of hydrogen lines. Theoretical upper limits calculated using the overshoot model of \citet{cunningham19} shown in blue, and limits using 1D convection \citep{koester09} are shown in red dashed. Shaded regions correspond to H abundances that could be explained by convective mixing in either the 3D overshoot (blue) or no overshoot (red) models. All observed H-abundances are well outside the range predicted by these models.}
 \label{fg:H-limit}
\end{figure}

The origin of hydrogen in DB(A) white dwarfs above 9000~K \citep{rolland18} can be reviewed in light of our improved description of the scenario of convective mixing. In Fig.~\ref{fg:H-limit} we plot in black circles the observed surface hydrogen abundance in a sample of 79 He-rich white dwarfs from \citet{rolland18}. Open circles correspond to upper limits inferred from a non-detection of hydrogen.

For each effective temperature, the convective mixing scenario predicts a mass of hydrogen that gets mixed with the underlying helium layer. Since this mass increases with decreasing $T_{\rm eff}$, at any given temperature, the maximum amount of hydrogen possible in a DB(A) white dwarf is for an object that has just experienced mixing. Lower hydrogen abundances are possible for DB(A) stars that have mixed at higher temperatures. We restrict our discussion to this upper limit of hydrogen in DB(A) stars.

For the mass of the helium convection zone we use the results of a grid of 3D simulations of DBA white dwarfs \citep{elena2}. These simulations do not include a parameterisation for (helium) convective overshoot since DB(A) stars in that temperature range have deep convection zones for which a direct simulation is not yet possible. Including overshoot would result in larger helium convection zones, and therefore smaller upper limits on the surface abundance of hydrogen. This grid is interpolated iteratively so that the final surface hydrogen abundance is taken into account for the size of the convection zone.

The red, dashed line in Fig.~\ref{fg:H-limit} shows the predicted abundance of hydrogen in a He-rich white dwarf after convective mixing has occurred assuming a 1D mixed mass. The blue line shows the same quantity with the inclusion of the larger mixed mass from convective overshoot \citep{cunningham19}. In both cases the calculation assumes that the hydrogen is homogeneously mixed into a larger He-rich convection zone.

The shaded regions beneath these upper limits can be considered the region in which all observations should lie if explained only by convective mixing. We note that the scenario of convective dilution predicts even smaller hydrogen abundances \citep{genest-Beaulieu19}. The entire observed sample of \citet{rolland18} lies well outside of these regions, with hydrogen abundances being $\approx$ 3--5~dex higher than the upper limit derived from 3D overshoot results. It seems clear that another mechanism must be invoked to explain these hydrogen abundances. 3D effects are clearly not able to provide a better fit to the observed hydrogen abundances. We favour the accretion of planetesimals as the most likely source of the observed hydrogen. 

In principle, hydrogen accretion could cause a reverse change from He- to H-dominated atmospheres. However, recent studies on the spectral evolution of helium atmosphere white dwarfs and their accretion of hydrogen have suggested that this is an extremely rare scenario \citep{ngf17,rolland18}. As such our assumption that the relative number of helium-atmosphere white dwarfs should increase monotonically with cooling age is likely a robust one.

\section{Conclusions}
\label{conclusions}
We have presented a statistically significant (>5$\sigma$) detection of white dwarfs undergoing a transition from hydrogen-dominated to helium-dominated atmospheres as they cool across the effective temperature range 20\,000--9000~K. This was done using the largest volume limited sample (133~pc) of white dwarfs for any previous study of this kind, with the precise determinations of effective temperature utilising \gaia\ photometry. We have characterised the temperature dependence of the rate of spectral evolution and used the most current grid of convection zone sizes for DA white dwarfs to determine the distribution of total hydrogen masses in white dwarfs. We find that the observed distribution of hydrogen shells in the white dwarf population peaks in the range of $\log M_{\mathrm{H}}/M_{\rm WD}$ from $-10$ to $-4$, with 60\% of all objects found within that range. Another 25\% of white dwarfs have thin hydrogen masses in the range $-14 < \log M_{\mathrm{H}}/M_{\rm WD} < -10$, and finally 15\% have even thinner hydrogen masses ($\log M_{\mathrm{H}}/M_{\rm WD}$ < $-14$).

These results have implications for models of pulsations in white dwarfs, stellar evolution - in particular during the AGB - and the accretion of material after the formation of white dwarfs. In the future, volume-complete spectroscopic samples will be able to increase the range in mass for which we can constrain the total amount of hydrogen in a white dwarf.

\section*{Acknowledgements}
This work has made use of data from the European Space Agency (ESA) mission {\it Gaia} (\url{https://www.cosmos.esa.int/gaia}), processed by the {\it Gaia}
Data Processing and Analysis Consortium (DPAC, \url{https://www.cosmos.esa.int/web/gaia/dpac/consortium}). Funding for the DPAC has been provided by national institutions, in particular the institutions participating in the {\it Gaia} Multilateral Agreement.

Funding for SDSS-III has been provided by the Alfred P. Sloan Foundation, the Participating Institutions, the National Science Foundation, and the U.S. Department of Energy Office of Science. The SDSS-III web site is http://www.sdss3.org/.

SDSS-III is managed by the Astrophysical Research Consortium for the Participating Institutions of the SDSS-III Collaboration including the University of Arizona, the Brazilian Participation Group, Brookhaven National Laboratory, Carnegie Mellon University, University of Florida, the French Participation Group, the German Participation Group, Harvard University, the Instituto de Astrofisica de Canarias, the Michigan State/Notre Dame/JINA Participation Group, Johns Hopkins University, Lawrence Berkeley National Laboratory, Max Planck Institute for Astrophysics, Max Planck Institute for Extraterrestrial Physics, New Mexico State University, New York University, Ohio State University, Pennsylvania State University, University of Portsmouth, Princeton University, the Spanish Participation Group, University of Tokyo, University of Utah, Vanderbilt University, University of Virginia, University of Washington, and Yale University.

The research leading to these results has received funding from the European Research Council under the European Union's Horizon 2020 research and innovation programme n. 677706 (WD3D).




\bibliographystyle{mnras}
\bibliography{mybib} 






\appendix


\bsp	
\label{lastpage}
\end{document}